\documentclass[aps,prb,superscriptaddress,reprint,longbibliography,amsmath,amssymb]{revtex4-2}

\usepackage{natbib}
\usepackage{ulem}

\usepackage{appendix}

\usepackage{graphicx}
\usepackage{xcolor}

\usepackage{bm}

\newcommand*{\lc}[1]{\textcolor{black}{#1}}
\newcommand*{\al}[1]{\textcolor{black}{#1}}

\begin{document}

\title{Crossover from renormalized 
to conventional diffusion near the 3D Anderson localization transition for light}

\author{Laura A. Cobus}
\email{laura.cobus@auckland.ac.nz}
\affiliation{Institut Langevin, ESPCI Paris, CNRS, PSL University, 1 rue Jussieu, F-75005 Paris, France}
\affiliation{Dodd-Walls Centre for Photonic and Quantum Technologies, New Zealand
and Department of Physics, University of Auckland, Private Bag 92109, Auckland, New Zealand}
\author{Georg Maret}
\affiliation{Fachbereich Physik, Universit\"{a}t Konstanz, D-78457 Konstanz, Germany}
\author{Alexandre Aubry}
\email{alexandre.aubry@espci.fr}
\affiliation{Institut Langevin, ESPCI Paris, CNRS, PSL University, 1 rue Jussieu, F-75005 Paris, France}

\date{\today}

\begin{abstract}

We report on anomalous light transport in the strong scattering regime.  Using low-coherence interferometry, we measure the reflection matrix of titanium dioxide powders, revealing crucial features of strong optical scattering which can not be observed with transmission measurements:  (i) a subdiffusive regime of transport at early times of flight that is a direct consequence of predominant recurrent scattering loops, and (ii) a 
\lc{crossover} to a conventional, but  extremely  slow,  diffusive  regime  at  long  times. These  observations  support  previous predictions that near-field coupling between scatterers prohibits Anderson localization of light in three-dimensional disordered media.
\end{abstract}

\maketitle

Since its discovery in 1958~\cite{Anderson1958}, Anderson localization (AL) has been the subject of intense research and debate. This unusual phenomenon can be described as the suppression or halting 
of wave propagation, arising solely from wave interference effects caused by disorder~\cite{Anderson1958,John1984,John1991}.  
In three-dimensions, wave transport exhibits a true phase transition~\cite{Evers2008} from conventional diffusion to AL, which occurs for some critical amount of disorder/energy, or alternately, at some critical time for a system in which waves  explore the disordered system over time. The requirement that waves must interact with a critical amount of disorder makes the experimental observation of localization in 3D notoriously difficult.

First predicted for quantum particles~\cite{Anderson1958}, AL was later extended theoretically for classical waves~\cite{John1984,John1991}. Light was proposed as a good candidate with which to study localization~\cite{John1991}, as photons are free from the inter-particle interactions which complicated early experiments with electrons~\cite{Rosenbaum1980,Lee1985}. In 1997, Wiersma et al. claimed the first experimental observation of 3D AL for classical waves, studying the thickness-dependent scaling of the optical transmission coefficient in white paint powders~\cite{Wiersma1997}. In 2006, Aegerter et al. observed anomalously slow decays of transmitted optical intensity~\cite{Aegerter2006} consistent with predictions for 3D localization. Similar subsequent reports studied the spatio-temporal behaviour of transmitted intensity~\cite{StorzerPhD2006,Sperling2013}. However, later reinterpretation of all of these experiments found that the results could also be explained  either by absorption~\cite{Scheffold1999,VanDerBeek2012,Sperling2016} or fluorescence~\cite{Sperling2016}, both of which can mimic signatures of localization. Meanwhile, AL in 3D was unambiguously observed using acoustic vibrations in elastic networks
~\cite{Hu2008,Cobus2016} and with cold atoms in random potentials
~\cite{Chabe2008,Kondov2011,Jendrzejewski2012}.

The ongoing lack of evidence for 3D localization of light motivated a resurgence of theoretical work in search of explanations. Some argued that none of the materials tested so far scatter light strongly enough to achieve localization
~\cite{Sperling2016,Skipetrov2016,Cottier2019}. Others proposed that the onset of localization is prevented by the dipole-dipole interactions between close-packed scatterers~\cite{John1992,Skipetrov2014,Bellando2014,Naraghi2015,Naraghi2016,Escalante2017,Tiggelen2021} which are inherent to the vector nature of light~\cite{Skipetrov2014,Cherroret2016}. This is in fact a relatively old idea~\cite{John1992}, recently revived by Skipetrov and Sokolov to predict that 3D localization should exist for (vector) elastic waves, but not necessarily for light~\cite{Skipetrov2014,Skipetrov2019}. Around the same time, Naraghi et al. proposed a theory in which localization and near-field coupling are modeled as competing effects~\cite{Naraghi2015,Naraghi2016}. The model predicts multiple regimes of transport as the waves explore the medium. In particular, this includes a 
\lc{crossover} from a 
\lc{subdiffusive/critical regime} to conventional diffusion, which 
is brought about when near-field effects induce an opposing energy `leak' which destroys localization. Supporting this picture are experimental measurements of the path length distribution of the optical energy flux reflected from white powder~\cite{Naraghi2016}. These measurements, however, are not necessarily independent of absorption or fluorescence, and can not indicate whether or not localization is attained prior to the 
\lc{crossover}.

\begin{figure*}[t]
	\centering\centering\includegraphics[width=\textwidth]{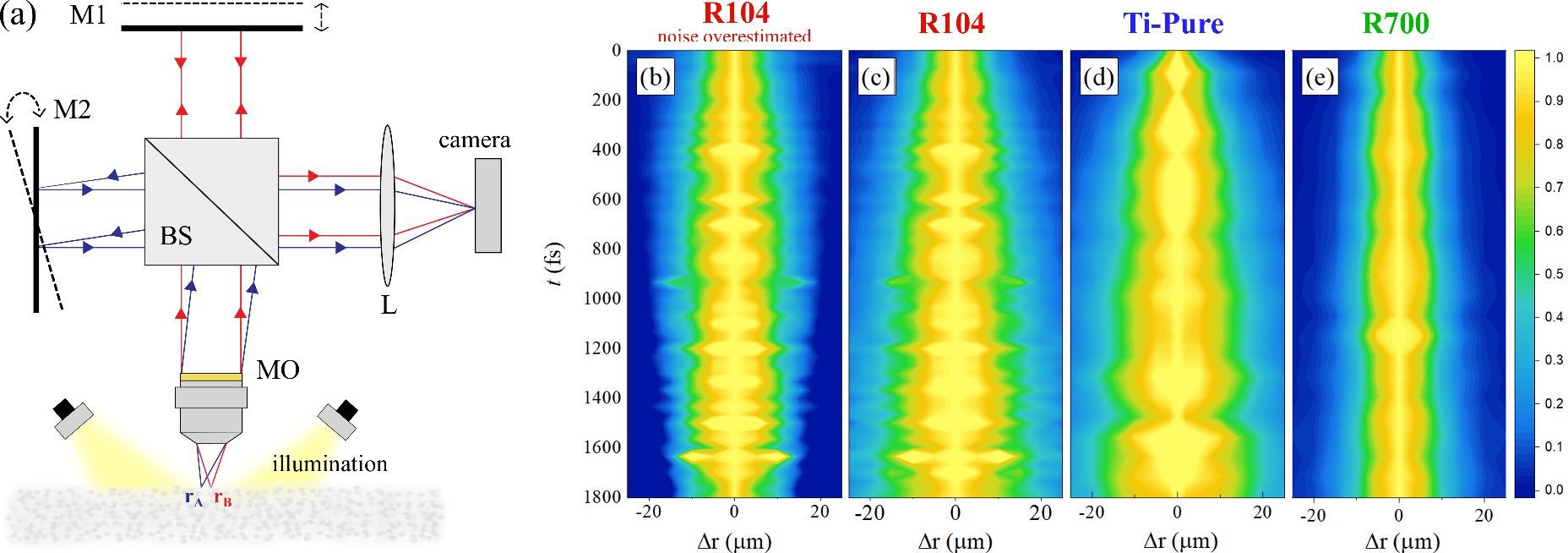}
	\caption{(a) Low-coherence interferometry measures the cross-correlation between points $\mathbf{r_A}$ and $\mathbf{r_B}$ at the sample surface~\cite{Badon2015}
	(MO: microscope objective, \lc{BS: beam-splitter}). 
(b,c) Comparison of intensity extraction approaches for sample R104. Diffuse halos for R104 are shown; in (b), the noise subtraction method overestimates the noise level, resulting in a narrower halo at long times than does the convergence method shown in (c).	$I(\Delta r,t)/I(0,t)$ (extracted via the convergence method) are also shown for samples (d) Ti-Pure and (e) R700. }
	\label{fig:setup}
\end{figure*}

Now, we report experimental measurements which confirm the existence of this predicted 
\lc{crossover}, and which indicate that the first regime is 
\lc{most likely not localized.} 
We use a low-coherence interferometry technique which is inspired by previous
studies in acoustics~\cite{Weaver2001,Derode2003} and seismology~\cite{Campillo2003,Larose2006}. Our approach enables the measurement of time-dependent Green's functions between points at the surface of a medium which is illuminated by an incoherent light source~\cite{Badon2015}. With this information, the full spatio-temporal spreading of wave energy can be studied~\cite{Badon2016}. We modify this technique to examine some of the strongest scattering samples which exist for light -- titanium dioxide (TiO$_2$) powders. Importantly, our  approach ensures that neither absorption~\cite{Page1995,Hu2008} nor fluorescence contribute to the experimental results. Using the model of Naraghi et al. as a starting point, we derive predictions for our observables. The agreement of our results with this theory provides the strongest experimental evidence to date that AL is suppressed for light propagating in 3D random media. 

\section{Measuring the reflection matrix}

Samples investigated in this work were three different types of TiO\textsubscript{2} powder: a pure anatase phase (Ti-Pure) obtained from Aldrich, and two types of rutile phase powders (R104 and R700) which are commercially available from DuPont as pigments for white paint. \lc{Significant Mie scattering resonances can be achieved in these powders, as mean particle size $\langle d\rangle\sim 400-800$~nm~\cite{StorzerPhD2006,BuhrerPhD2012} is on the order of the wavelength of the illuminating light.} The powders were compressed into pastille form to decrease the transport mean free path, $\ell^*$, and hence increase optical scattering. Scattering strength can be characterized by  $k_{0}\ell^*$, the product of optical wave number $k_{0}$ in vacuum and $\ell^*$. A value of $k_0\ell^*\sim 1$ indicates very strong scattering, and has been used as an approximate criterion for AL~\cite{Abrahams1979}. Previous measurements have reported $k_{0}\ell^*\sim 5-6$, $3-4$ and $2-3$ for compressed Ti-Pure, R104 and R700 respectively~\cite{SperlingPhD2015,Schertel2019}, indicating that light experiences very strong scattering in all three powders. 
Absorption, on the other hand, is relatively weak, as absorption time $\tau_a$ is on the order of $1$~ns~\cite{SperlingPhD2015}.

We measure the spatio-temporal transport of light in 
\lc{the white-paint samples} using the low-coherence interferometry apparatus  introduced by Badon \textit{et al.}~\cite{Badon2015,Badon2016} (Fig.\ \ref{fig:setup}a). A low-coherence broadband light source ($650-850$~nm, radiant flux $\sim5\times10^5$ W.cm$^{-2}$, coherence time $\tau_c=10$~fs~\cite{Badon2015}) isotropically illuminates the sample surface. 
The backscattered light is collected by a microscope objective (NA=0.25) and sent to a Michelson interferometer, which here is used as a spatio-temporal field correlator. 
A CCD camera conjugated with the sample surface records the output intensity: 
\begin{equation}
S_{\alpha}(\mathbf{r},\mathbf{r}+\Delta \mathbf{r},t) =  \int_{0}^T |e^{\alpha}\psi(\mathbf{r},t+\tau)+\psi(\mathbf{r}+\Delta \mathbf{r},\tau)|^2  \mathrm{d}\tau , 
\label{eq:S}
\end{equation}
where $\tau$ is the absolute time, $\mathbf{r}$ the position vector on the CCD  screen, $\psi(\mathbf{r},\tau)$ the scattered wave field associated with the first interference arm, $T$ the integration time of the camera, and $\alpha$ an additional phase term controlled with a piezoelectric actuator placed on mirror $M_1$. The tilt of mirror $M_2$ allows a displacement $\Delta \mathbf{r}$ of the associated wave-field at the  camera. The motorized translation of mirror $M_1$ induces a time delay $t = \delta/c$ between the two interferometer arms, with $\delta$ the optical path difference (OPD) and $c_0$ the light celerity in vacuum. The interference term is extracted from the four intensity patterns (Eq.\ \ref{eq:S}) recorded at $\alpha=0,$ $\pi/2$, $3\pi/2$ and $\pi$ (``four phase method''~\cite{Badon2016}). On each pixel of the  camera, we thus measure the cross-correlation $C(\mathbf{r}_A,\mathbf{r}_B,t)$ between scattered wave-fields, $\psi_c(\mathbf{r}_A,\tau)$ and $\psi_c(\mathbf{r}_B,\tau)$, associated with each arm of the {interferometer}:
\begin{equation}
\label{cor}
C_T(\mathbf{r}_A,\mathbf{r}_B,t)  = \frac{1}{T} \int_{0}^T \psi(\mathbf{r}_A,t+\tau)\psi^*(\mathbf{r}_B,\tau)  \mathrm{d} \tau ,
\end{equation}
{$\mathbf{r}_A$ and $\mathbf{r}_B$} for an infinite integration time~\cite{Badon2015,Badon2016}. In practice, the limited numerical aperture of our experimental device induces aberrations. This implies that our measurement scheme does not give  direct access to the true Green's function between points $\mathbf{r}_A$ and $\mathbf{r}_B$; rather, we measure the response between a virtual source at point $\mathbf{r}_A$ and a virtual detector at point $\mathbf{r}_B$ \al{(see Appendix \ref{sec:Theory0})}. The characteristic size of these sources is governed by the resolution length $\delta r$ of the imaging system, here equal to $1.8$ $\mu$m. 
This matrix contains the impulse responses between points at the sample surface, $\mathbf{r}_A$ and $\mathbf{r}_B$~\cite{Badon2016}.
The spatio-temporal behaviour of the wave energy density at the surface of the sample is described by the ensemble average of the impulse response intensity: 
$I(\Delta r,t) \equiv \langle \left| R (\mathbf{r}_A,\mathbf{r}_B,t)  \right |^2 \rangle $, with $\Delta r=|\mathbf{r}_A-\mathbf{r}_B|$. In practice, this ensemble average is obtained via a spatial average over pairs of points $\mathbf{r}_A$ and $\mathbf{r}_B$ separated by the same distance $\Delta r$.


\section{Overcoming shot noise and fluorescence}
\label{sec:conv_curve_fitting}

Our passive imaging method enables the extraction of $I(\Delta r,t)$ from fluorescence and noise contributions that usually pollute active measurements~\cite{Sperling2013,Sperling2016}. Here, we discuss how this noise contributes to the measured signals, and report a new method to cleanly extract the desired signals from noise.

Because of the incoherence of the illumination, the ensemble average of $C(\mathbf{r_A},\mathbf{r_B},t)$ should be theoretically achieved by integrating the interferometric signal over an infinite integration time $T$ (Eq.\ref{cor}):
\begin{equation}
\lim\limits_{T \rightarrow +\infty} C_T(\mathbf{r_A},\mathbf{r_B},t) =  R(\mathbf{r_A},\mathbf{r_B},t) .
\end{equation}
In practice, $T$ is finite and the convergence of $C_T$ towards its ensemble average cannot be completely assured. The situation is made worse by the fact that, for large $\Delta {r}$ and time lapse $t$, the signal of interest can be very weak, and undesirable {noisy} contributions such as shot noise and fluorescence may dominate.

To evaluate the convergence of $C_T$ towards $R$, one can consider the intensity profile $I_T(\Delta r,t)$ averaged over pairs of points $\mathbf{r}_A$ and $\mathbf{r}_B$ separated by the same distance $\Delta r$:
\begin{equation}
    I_T(\Delta r,t)=\langle |C_T(\mathbf{r}_A,\mathbf{r}_B,t)|^2 \rangle_{\lbrace (\mathbf{r}_A,\mathbf{r}_B) \, | \, | \mathbf{r}_A -\mathbf{r}_B|=\Delta r \rbrace }.
\end{equation}
Here, we have assumed that disorder is statistically homogeneous. The evolution of $I_T(\Delta r)$ with $T$ is given by~\cite{Badon2016}
\begin{align}
\label{eq:convergence}
I_T \left(\Delta {r},t\right) 
&=  I \left(\Delta {r},t\right) 
\nonumber \\
& + \frac{\delta t}{T}\langle \left|N(\Delta r,t)\right|^2\rangle_{\lbrace (\mathbf{r}_A,\mathbf{r}_B) \, | \, | \mathbf{r}_A -\mathbf{r}_B|=\Delta r \rbrace } ,
\end{align}
where
\begin{align}
\label{eq:I}
   I(\Delta r,t)= \langle | R(\mathbf{r_A},\mathbf{r_B},t) |^2 \rangle_{\lbrace (\mathbf{r}_A,\mathbf{r}_B,t) \, | \, | \mathbf{r}_A -\mathbf{r}_B|=\Delta r \rbrace }
\end{align} 
is the mean intensity profile for an infinite integration time. 
$N(\Delta r,t)$ represents the contribution of incoherent noise whose coherence time is governed by the source bandwidth and thus scales as $\delta t/T$. The noise contribution corresponds to the part of the wave-field whose cross-correlation function vanishes with the average over $T$ in Eq.~\ref{eq:convergence}. Shot noise and fluorescence signals that result from spontaneous emission events thus emerge along this noise contribution.

\begin{figure}[h]
	\centering\centering\includegraphics[width=\columnwidth]{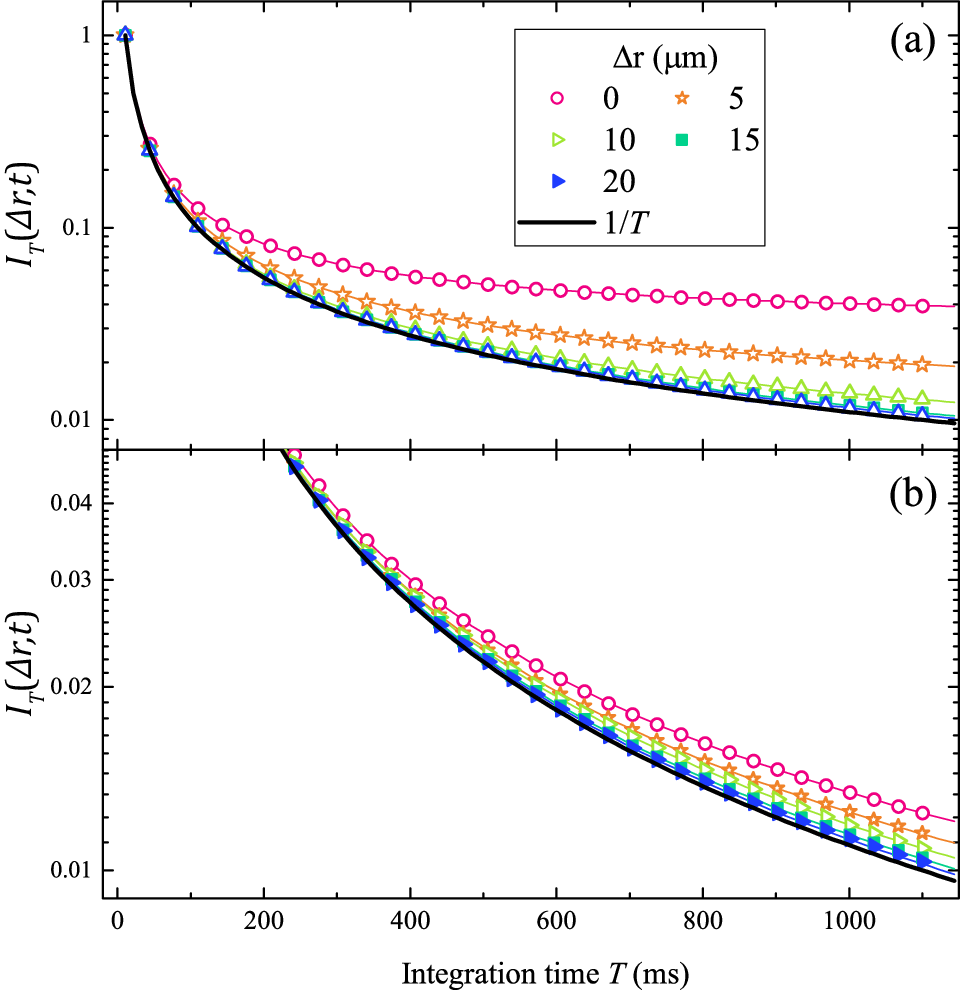}
	\caption{Average correlation function (solid symbols) as a function of integration time (number of averages). Data shown is for R104 for several representative values of $\Delta {r}$, at times of flight (a) $t=33$~fs and (b) 1000 fs. Corresponding fits [Eq.\ (\ref{eq:convergence})] are shown as solid lines. In both (a) and (b) $1/T$ is represented by a thick black solid line. For clarity, experimental uncertainties are not shown. }
	\label{ConvCurves_fig}
\end{figure}

Figure\ \ref{ConvCurves_fig} shows representative results for $I_T \left(\Delta r,t\right) $ as a function of $T$ (solid symbols). Results for several values of {$\Delta {r}$} are shown for times of flight (a) $t=0$~fs and (b) $t=1200$~fs. 
For small integration times, {$I_T \left(\Delta r,t\right) $} decreases as $1/T$, as expected for noise. At large $T$, and for small values of $\Delta r$ and $t$, $I_T \left(\Delta r,t\right) $ plateaus at the value of {$I \left(\Delta r,t\right)$}.
In previous works~\cite{Badon2015,Badon2016}, `by eye' examination of such convergence curves was performed to estimate whether or not {$I_T \left(\Delta r,t\right) $} had converged satisfactorily towards {$I \left(\Delta r,t\right)$}. Points for which the convergence curve is well above the noise level (black dotted lines in Fig.\ \ref{ConvCurves_fig}) were deemed acceptable. 
This noise level was then taken to be the intensity measured at the maximum $\Delta r$ and $t$ of the scan, assumed constant over space and time, and subtracted from {$I_T \left(\Delta r,t\right)$} to obtain {$I_T \left(\Delta r,t\right)$}. 
For accurate measurements at much longer times of flight, however, this method is no longer viable. Figure\ \ref{ConvCurves_fig} shows an example of convergence curves measured for sample R104. Even with a very large integration time, {$I_T \left(\Delta r,t\right)$} does not converge to a constant value for some values of {$\Delta r$} at the relatively short time of $t=33$~fs [Fig.\ \ref{ConvCurves_fig}(a)], or for any spatial position at the later time $t=1000$~fs [Fig.\ \ref{ConvCurves_fig}(b)]. It is clear that another method of extracting $I \left(\Delta r,t\right)$ from noise is required.
To this end, we introduce a simple yet powerful alternate method. We fit the experimental {$I_T \left(\Delta r,t\right)$} versus $1/T$ with a straight line; the slope gives the noise level while the y-offset gives {$I \left(\Delta r,t\right)$ according to Eq.~\ref{eq:convergence}}. This fit is a weighted fit, with experimental uncertainty {$\sigma(T)$ taken to be the error in the mean over pixels $\mathbf{r}$ in the calculation of 
$I_T(\Delta r,t)$ (Eq.\ \ref{eq:I}), for each integration time $T$}. 
The uncertainty in $I \left(\Delta r,t\right)$ is given by~\cite{Bevington2002}
\begin{equation}
\sigma_I = \sqrt{\frac{1}{\Delta}\sum_T\frac{(1/T)^2}{\sigma(T)^2}} ,
\end{equation}
where
\begin{equation}
    \Delta=\sum_T\frac{1}{\sigma(T)^2}\sum_T\frac{(1/T)^2}{\sigma^2(T)}-\left(\sum_T\frac{(1/T)^2}{\sigma^2(T)} \right)^2 .
\end{equation}
In Fig.\ \ref{ConvCurves_fig}, representative results for this fitting procedure are shown (solid lines). For all positions and times, the data can be very well fit with the form of Eq.\ (\ref{eq:convergence}). 

Figures \ref{fig:setup}(b,c) show normalized intensity profiles for sample R104, calculated using (b) the noise subtraction method from previous studies~\cite{Badon2015,Badon2016}, and (c) the {$I_T \left(\Delta r,t\right)$} fitting approach introduced in this work. 
In Fig.\ \ref{fig:setup}(b), the data give the impression that energy is spreading much more slowly than it appears in Fig.\ \ref{fig:setup}{(c}).
This spread even seems to stop at long times of flight -- a feature that could be wrongly attributed to Anderson localization~\cite{Sperling2013,Sperling2016}. Such contributions from incoherent noise can thus hamper accurate observations of the diffuse halo, as had already been observed for fluorescent noise in a transmission geometry~\cite{Sperling2013,Sperling2016}.  The `false' plateau reached by the diffuse halo in Fig.~\ref{fig:setup}{(b)} occurs because the noise subtraction method overestimates the noise level at the largest $\Delta r$ and $t$ in the measured range, for a (finite) number of averages. Subtracting this estimated noise level then results in narrower intensity profiles at long times. In contrast, the $I_T \left(\Delta r,t\right)$ fitting approach uses the data gathered over a finite range of {$\Delta r$} and $t$ to properly estimate the noise level at each point (${\Delta {r}}$,$t$).


\section{Quantifying the energy spread}

To eliminate the effect of absorption, we normalize the measured intensity by $I(\Delta r=0,t)$~\cite{Page1995}. 
Figures\ \ref{fig:setup}(d,e) show the resulting normalized intensity profile,  $I(\Delta r,t)/I(0,t)$, for samples Ti-Pure and R700. The difference between samples is immediately obvious: energy spreads more slowly in R700, the more strongly-scattering sample. To quantify this effect,  we 
\lc{present} theoretical predictions for $I(\Delta r,t)$ with which to compare experimental data.

In the 
\lc{diffusive} regime, $I(\Delta r,t)$ can be expressed as the sum of two components. The first contribution is the incoherent average of the intensity of each individual scattering path, $I_\text{inc}$. In real space, $I_\text{inc}$ describes the spatio-temporal spreading of the wave energy density inside the sample -- the so-called \textit{diffuse halo}. This spreading can be directly quantified by measuring $w(t)$, the \textit{transverse width} of $I_\text{inc}(\Delta r,t)$~\cite{Page1995,Aubry2007}. $I_\text{inc}(\Delta r,t)$ can be expressed as follows~\cite{Patterson1989} \lc{(see Appendix\ \ref{sec:TheoryI} for a full derivation)}:
\begin{equation}
\label{eq:trans_int_ratio_w2}
I_\text{inc}(\Delta r,t) = {P_R(t)}\exp{\left[ -\Delta r^2/w^2(t) \right]},
\end{equation}
with $P_R(t)=(c e^{-t/\tau_a})/[(2\pi^{3/2}) w^3(t)]$, the probability of return to the origin, and
\begin{equation}
 \label{eq:w2_4DBt}
 w^2(t) = 4 D_B t.    
\end{equation} 
$D_B$ is the Boltzmann diffusion coefficient~\cite{Patterson1989,Page1995}{, $\tau_a$ is the absorption time, and $c$ the speed of light in the sample}. For anomalous wave transport, $w^2(t)$ no longer exhibits a linear increase with time~\cite{Hu2008,Lemarie2010,Cherroret2010}. 

\begin{figure}
	\centering
	\includegraphics[width=\columnwidth]{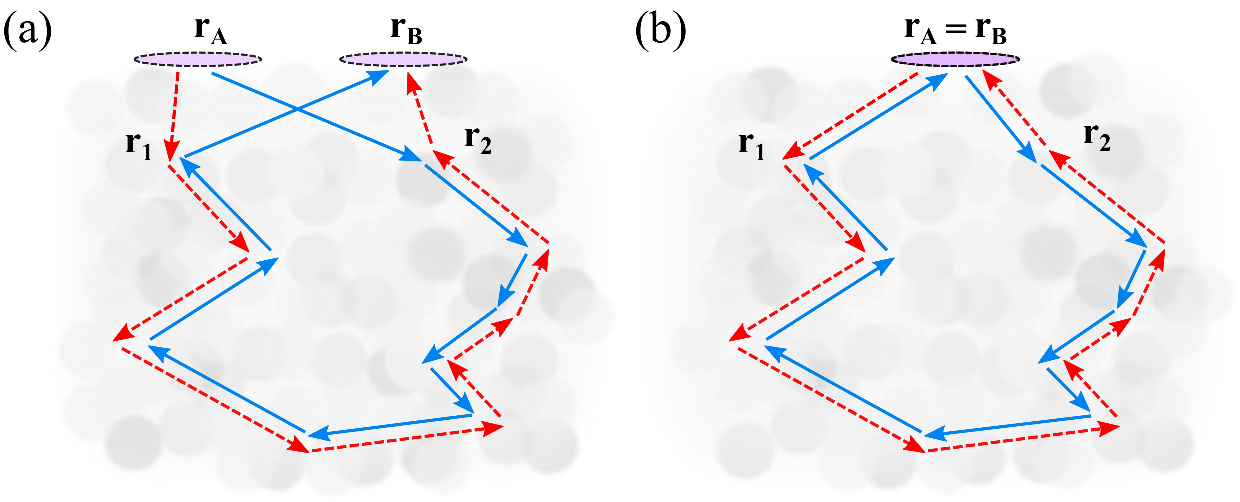}
	\caption{(a) Coherent backscattering arises from interference between reciprocal paths. Pink oblongs represent the size of virtual sources/receivers at $\mathbf{r}_A$ and $\mathbf{r}_B$. (b) When source and receiver coincide, constructive interference is maximized.	}
	\label{fig:MS_paths}
\end{figure}
The second contribution to $I(\Delta r,t)$ is the coherent intensity $I_\text{coh}(\Delta r,t)=I_\text{coh}(0,t)F(\Delta r)$. This factor \lc{is a correction to the diffusion approximation to account} 
for \textit{weak localization}~\cite{Albada1985,Wolf1985} in which waves travelling along pairs of reciprocal paths in opposite directions undergo constructive interference (Fig.\ \ref{fig:MS_paths}). 
\lc{Experimentally, the effect of weak localization can be observed as} \textit{coherent backscattering} (CBS)~\cite{Albada1985,Wolf1985} -- a peak-shaped enhancement $F(\Delta r)$ in the spatial distribution of average backscattered intensity. The CBS peak is maximum at $\Delta r=0$, and its enhancement factor, $A$, can be defined by the relation $I_\text{coh}(0,t)=(A-1)I_\text{inc}(0,t)$. While in $\mathbf{k}$-space this CBS peak narrows as time increases~\cite{Vreeker1988,Tourin1997,Jendrzejewski2012,Hainaut2017}, in real-space its shape is stationary~\cite{Margerin2001,Larose2004,Aubry2007}. 
Ideally, for an experiment with point-like sources and detectors on the medium surface, this real-space CBS peak would have the form\al{~\cite{Margerin2001}}  
\lc{\begin{equation}
\label{eq:F_CBS}
    F(\Delta r)=[\sin({k\Delta r})/k\Delta r]^2 \exp({-\Delta r/\ell_s)}. 
\end{equation}
The width of the CBS peak} would then scale as $\lambda/2$ or $\ell_s$, the scattering mean free path. \al{The enhancement factor ranges theoretically from $A=2$ in the diffusive regime to $A=3$ in the localized regime~\cite{Cherroret2021}.}  
Here, however, the CBS peak shape is dictated by the impulse response $H$ of our imaging system
such that the peak is widened \al{(see Appendix\ \ref{sec:TheoryI} for a full derivation): 
\begin{equation}
    F(\Delta \mathbf{r}) \propto |H {\ast} H|^2  (\Delta \mathbf{r}),
\end{equation}} 
and the enhancement factor is reduced: 
\lc{\begin{equation}
    A = 1+ \beta |H {\ast} H|^2  (\Delta \mathbf{r}=\mathbf{0}) / [|H|^2 \circledast |H|^2](\Delta \mathbf{r}=\mathbf{0}),
\end{equation}
\al{with the factor $\beta$ ranging from 1 to 2 (from the diffusive to the localized regime). The}} symbols ${\ast}$ and ${\circledast}$ denote convolution and correlation products over $\Delta \mathbf{r}$, respectively.

Thus, for conventional diffusion, $I(\Delta r,t)$ has time-dependent diffuse halo such that \al{(see Appendix\ \ref{sec:TheoryI}):}
\begin{align}
\label{eq:intensity_profile_w2_and_CBS}
   \frac{ I(\Delta r,t)}{I(0,t)} = \frac{1}{A}e^{-\Delta r^2/w^2(t)}+\left(1-\frac{1}{A}\right)F(\Delta r).   
\end{align}
\begin{figure}
	\centering
\includegraphics[width=\columnwidth]{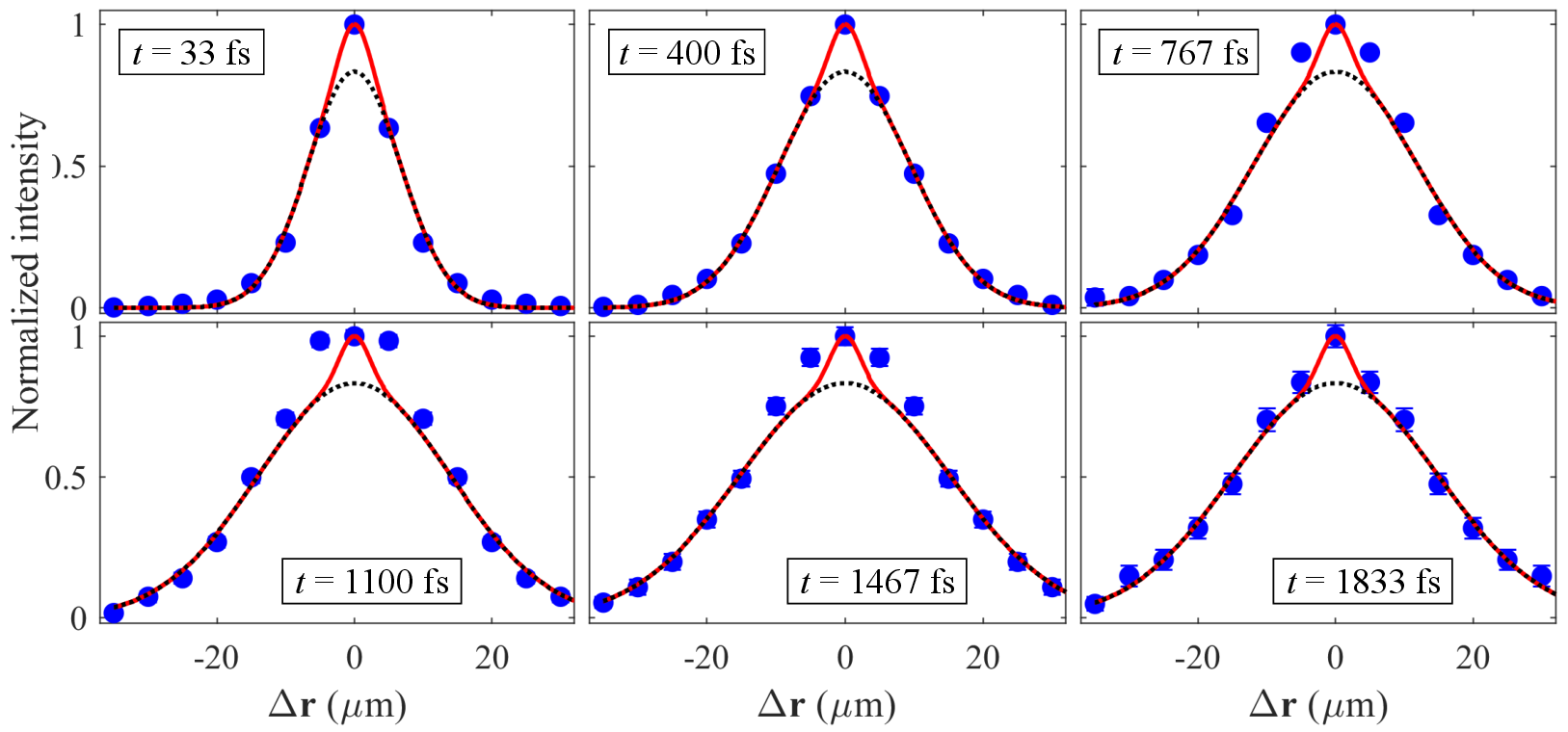}
	\caption{Normalized intensity profiles $I(\Delta r,t)/I(0,t)$ for sample R104 (solid symbols) at different times $(t)$. Note that experimental error bars are smaller than the symbol size. Red solid lines show  theoretical fits with Eq.\ \ref{eq:intensity_profile_w2_and_CBS}, while black dotted lines represent only the diffuse halo term of Eq.\ \ref{eq:intensity_profile_w2_and_CBS}.
	}
	\label{fig:profiles_R104_2Gaussfits}
\end{figure}
Figure \ref{fig:profiles_R104_2Gaussfits} shows $I(\Delta r,t)/I(0,t)$ for sample R104 for six times-of-flight $t$ spanning the entire measurement range. The spatio-temporal spreading of wave energy is clearly exhibited, as is a small and constant CBS enhancement around $\Delta r=0$.  To quantify 
the energy spread in each sample, the experimental $I(\Delta r,t)/I(0,t)$ was fit with Eq.\ \ref{eq:intensity_profile_w2_and_CBS}. Fit parameters were $w^2(t)$ (a free parameter for each time $t$), and $A$ (held constant over time). 
The fitting gives very small values for $A$ which are caused by aberration effects in the experimental setup \al{(see Appendix\ \ref{sec:TheoryI})}: $A=1.12$ for Ti-Pure, $A=1.2$ for R104, and $A=1.1$ for R700. The temporal fluctuations of the experimental CBS peak observed in Fig.\ \ref{fig:profiles_R104_2Gaussfits} may be explained by variations of the impulse response $H$ with optical path difference of the Michelson interferometer (Fig.\ \ref{fig:setup}).

It is important to note that, while the shape of $I_\textrm{inc}(\Delta r,t)$ is only strictly expected to be Gaussian in the diffusion approximation \lc{(see also Appendix\ \ref{sec:TheoryI})}, ultrasound experiments have shown that a Gaussian fit and its parameter $w^2(t)$ can give a good quantification of spatio-temporal energy spreading, even in the localized regime~\cite{Cobus2018}.
In the present case, $I_\text{inc}(\Delta r,t)$ is well-described by a Gaussian for the entire time range under investigation  \al{(see Fig.~\ref{fig:profiles_R104_2Gaussfits} and Appendix\ \ref{sec:fittingIwiththeories} for \al{more} details}). 

\section{Deviation from conventional diffusion}

\begin{figure}
	\centering\includegraphics[width=\columnwidth]{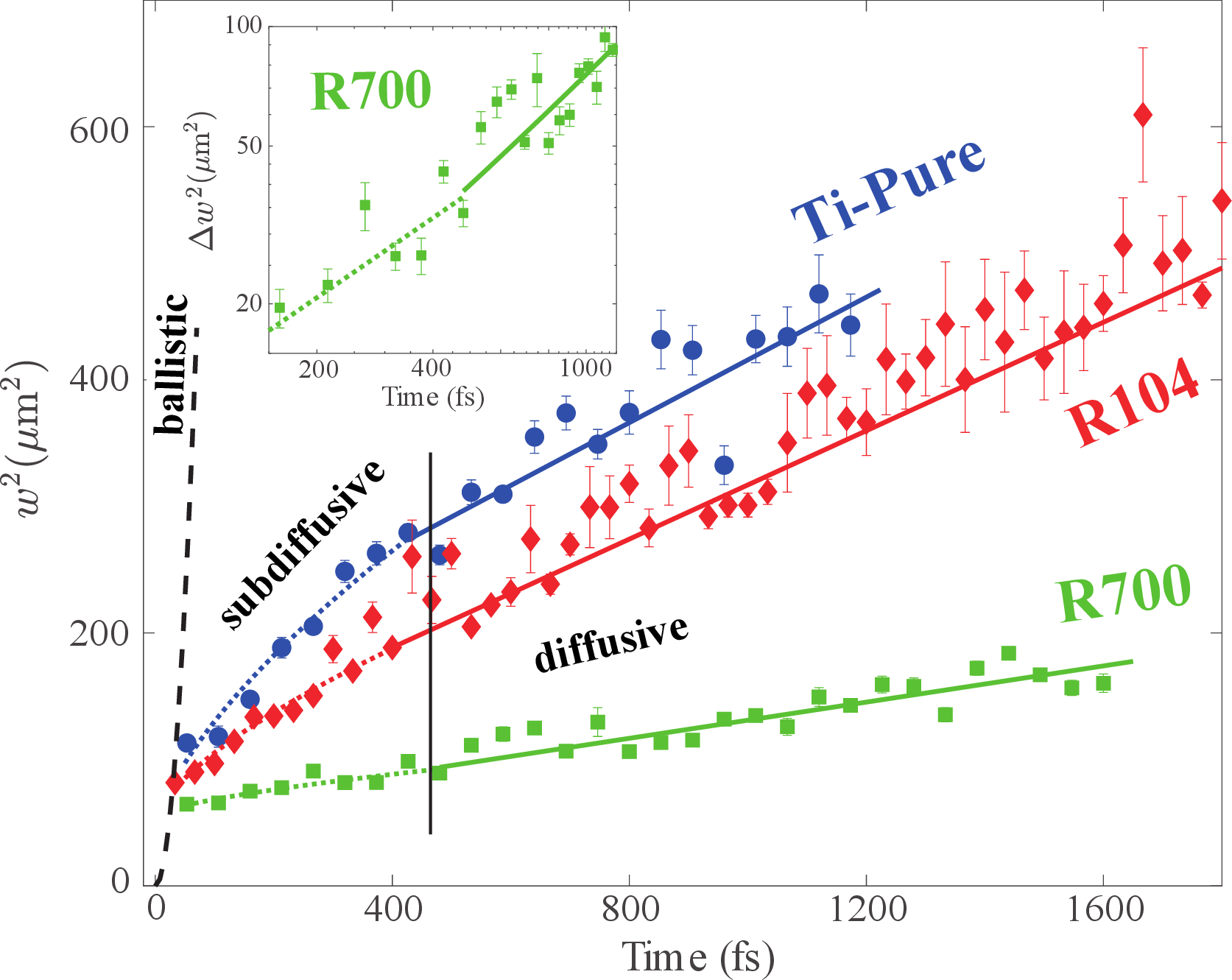}
	\caption{Transverse width {$w^2(t)$} for all three samples (symbols). Error bars represent the uncertainty in $w^2(t)$ due to the (weighted) 
	fitting of $I(\Delta r,t)/I(0,t)$ with Eq.\ \ref{eq:intensity_profile_w2_and_CBS}. Linear fits to the data (solid lines) give $D$ for each sample. In the \lc{subdiffusive} regime 
	($t< \tau_c$), the data is fit with Eq.\ \ref{w2_loc} (dotted lines). The ballistic light line  (dashed black line), indicates the lower limit for $w^2(t)$. Inset: 	$\Delta w^2$ through the subdiffusion-diffusion 
	\lc{crossover} for R700 (log-log scale). }
	\label{fig:w2}
\end{figure}
\lc{Figure\ \ref{fig:w2} shows experimental resutls for $w^2(t)$ (solid symbols). We observe that $w^2(t)$} 
does not agree with the diffusive prediction of Eq.\ \ref{eq:w2_4DBt}, especially at short times-of-flight ($t<500$ fs). This implies that our theoretical model must be altered to take into account the extreme strong scattering of our samples.

\subsection{Recurrent scattering and renormalized diffusion}

\begin{figure}[t]
	\centering\includegraphics[width=\columnwidth]{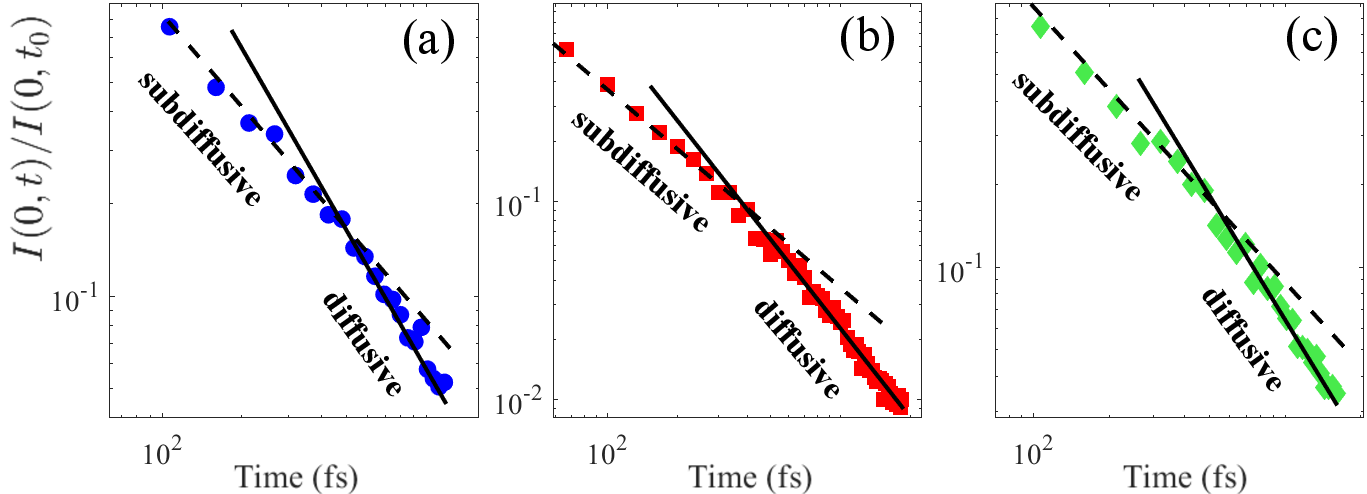}
	\caption{Return probability $P_R(t)$ normalized by its value at time $t_0 = 53$~fs for (a) Ti-Pure, (b) R104 and (c) R700 (symbols). Lines following $t^{-3/2}$ (solid, diffusive) and $t^{-1}$ (dashed, subdiffusive) are guides to the eye, not fits.}
	\label{fig:RS}
\end{figure}

Compared with conventional diffusion, a key feature of the strong scattering regime ($k\ell^*\sim 1$) is the predominance of recurrent scattering `loops', \textit{i.e} an increased probability for waves to pass nearby areas that they have previously visited~\cite{Vollhardt1982,Skipetrov2006,Aubry2014}. 
\lc{Observed in the near-field (on the sample surface), recurrent scattering is a subset of the path pairs which contribute to CBS (Fig.\ \ref{fig:MS_paths}b), being limited to path pairs in which source and receiver coincide.} The time-dependence of 
\lc{this} \textit{return probability} $P_R(t)$ can \lc{thus} be directly quantified in the reflection geometry by observing the back-scattered intensity at the source location\lc{,} since $I(0,t)=A \times P_R(t)$. 
Figure\ \ref{fig:RS} shows $I(0,t)$ for all three samples. As absorption is negligible for the time range of our measurements, $P_R(t)$ should scale as $t^{-3/2}$ in the diffuse regime~(Eq.~\ref{eq:trans_int_ratio_w2}). 

\lc{For stronger disorder ($k\ell^* \sim 1$), an increase in the 
return probability can cause a renormalization (decrease) of the diffusion coefficient. The self-consistent theory of localization predicts that, in 3D, {${D}$} scales as~\cite{Anderson1958,John1984}
\begin{equation}
   {{D}}\approx D_0 \ell^*\left(\frac{1}{\xi}+\frac{1}{L}+\frac{1}{L_A} \right),
    \label{eq:D_scalingtheory}
\end{equation}
where $D_0$ is the diffusion coefficient before rescaling, $\xi$ is the correlation length defining the spatial coherence of a wavefield in the delocalized regime, $L$ is the system size, and $L_A$ is the absorption length.} 
\lc{In reflection, the effective system size $L$ can be said to be the spatial extent of the 
wave energy $L(t) = \sqrt{6 D(t)t}$~\cite{Douglass2011}. Thus, before reaching localization ($L<<\xi$) and if absorption is negligible ($L << L_a$), Eq.~\ref{eq:D_scalingtheory} leads to the following time-dependence for the diffusion coefficient:
\begin{equation}
\label{locD}
    D(t) \simeq \frac{(D_0 \ell^*)^{2/3}}{(6t)^{1/3}}.
\end{equation}}
\al{This expression corresponds to a sub-diffusive regime in which}
\lc{the diffusion coefficient is successively renormalized as the propagating waves undergo recurrent scattering events -- a process which goes hand in hand with an increase in the return probability \al{and 
\lc{which, in principle, eventually results in} Anderson localization~\cite{Vollhardt1982,Skipetrov2006}}.} \lc{When such \textit{renormalized diffusion} occurs, both $I_\text{inc}$ (Eq.\ \ref{eq:trans_int_ratio_w2}) and $I_\text{coh}$ (Eq.\ \ref{eq:F_CBS}) will be influenced by the additional \al{interference effects}
arising from recurrent scattering. Experimentally, this influence can be observed in the behaviour of both $w^2(t)$~\cite{Cobus2018} and the return probability~\cite{Aubry2014}, as discussed in the next section.} 

\subsection{Subdiffusion-diffusion 
\lc{crossover}}

Replacing $D_B$ \lc{in Eq.~\ref{eq:trans_int_ratio_w2}} by the renormalized diffusion coefficient $D(t)$ 
gives the following scaling for 
\lc{the return probability in the 
\al{subdiffusive regime}}: $P_R(t)\propto t^{-1}$~\cite{footnote}. 
Comparison of 
\lc{our} theoretical predictions \lc{for $P_R(t)$} with the experimental data reveals the existence of a 
\lc{crossover} between two transport regimes at a 
\lc{characteristic} time
$\tau_{c}\sim 425$~fs for Ti-Pure, $\tau_{c}\sim 400$~fs for R104 and $\tau_{c}\sim 480$~fs for R700  (Fig.\ \ref{fig:RS}). 
Before $\tau_c$, $I(0,t)\propto t^{-1}$ which is characteristic of a regime 
\lc{in which the diffusion coefficient is continuously renormalized} (falling into the more general category of \textit{subdiffusion}). After  
the 
\lc{crossover} ($t>\tau_c$), $I(0,t)\propto t^{-3/2}$ as expected for diffusion.
Naraghi and Dogariu~\cite{Naraghi2016} 
have predicted such a 
\lc{crossover}, proposing that near-field coupling between scatterers constitutes a `leak' of energy from propagating paths (e.g. Fig.\ \ref{fig:MS_paths}) to evanescent channels.
This effect lessens the constructive interference created by recurrent scattering paths, preventing the localization of wave resonances. 
\lc{The crossover between the sub-diffusive and diffusive regime thus takes place when the probabilities of recurrent scattering and near-field leaking 
become comparable. The recurrent scattering probability, $p_{\times}$, is determined by the ratio between the trajectory volume, and the volume that the light explores inside the multiple scattering medium~\cite{Akkermans2007}. In a reflection geometry, $p_{\times}$ is stationary and given by~\cite{Naraghi2016}
\begin{equation}
\label{pcross}
    p_{\times}\sim \frac{\lambda^2}{2 \ell^{*2}}.
\end{equation}
For a high concentration of particles, energy can leak out of the diffusive channels because of the near-field interactions between scatterers which are less than a wavelength apart. The corresponding leakage probability, $p_{leak}$,
along a wave trajectory of length $s$ is given by~\cite{Naraghi2016}
\begin{equation}
\label{pleak}
    p_{leak} \sim \frac{3}{2} n_0 \sigma_{NF} s,
\end{equation}
where $n_0$ is the particle concentration and $\sigma_{NF}$ is the near-field cross-section of an individual scatterer. Setting $ p_{\times}$ (Eq.~\ref{pcross}) and $ p_{leak}$ (Eq.~\ref{pleak}) to be equal, a critical path length $s_c$ can be derived: 
\begin{equation}
    s_c \sim \frac{\lambda^2}{3 \ell^{*2} n_0 \sigma_{NF}} .
\end{equation}
Thus, the characteristic time can be theoretically expressed as
\begin{equation}
\label{eq:tau_c}
    \tau_c \sim \frac{\lambda^2}{3 c \ell^* } \frac{\sigma_t}{\sigma_{NF}},
\end{equation}
where $\sigma_t \sim 1/(n_0 \ell^*)$ is the transport cross-section of an individual scatterer. Using the values of $\ell^*\sim 0.3 $~$\mu$m~\cite{Schertel2019} and refractive index $n \sim 2.7$~\cite{SperlingPhD2015} measured in R700 at $\lambda_0=700$~nm, and taking $\tau_c\sim 480$~fs (Fig.\ \ref{fig:RS}c), we find $\rho \sim 10^{-2}$.
While this small value implies that the energy leak is negligible, it is for only a few scattering events; in the long-time limit the sheer number of these events results in the extinction of localization. 
We note that the observed values of $\tau_c$ do not scale with $\ell^*$. This is likely due to variation in $\rho$; although each sample was compressed with the same force, differences in particle size/shape could cause the volume fraction and $\rho$ to vary between samples. The differing chemical composition of the samples could also affect the near-field scattering cross-section.}\\

In the study of localization, the return probability $P_R(t)$ is valuable as it is a key quantity in the theoretical description of 
\lc{the process of renormalization of the diffusion coefficient}~\cite{Vollhardt1982,Skipetrov2006}. However, $P_R(t)$ is not necessarily free from absorption, and can exhibit surprising results in the critical regime due to boundary conditions and source shape~\cite{Aubry2014,Skipetrov2018}. It is thus unwise to rely on only $P_R(t)$ to distinguish between regimes. For this task, $w^2(t)$ is more reliable, being independent of all of the above issues~\cite{Page1995,Cobus2016,Cobus2018}, closely linked to $D$~\cite{Page1995,Patterson1989,Cobus2018}, and giving access to the very earliest times.

Figure\ \ref{fig:w2} shows experimental results for $w^2(t)$. At long times of flight ($t>\tau_c$), $w^2(t)$ increases linearly with time.
This behaviour is predicted by the diffusion approximation, in which $w^2(t)=4Dt+w_c^2$, where $w_c^2$ is the lateral extension of the energy halo at 
\lc{characteristic} time $\tau_c$. Linear fits to $w^2(t)$ give a direct measurement of $D$ for each sample: $D=62\pm 10$~m\textsuperscript{2}/s for Ti-Pure, $D=54\pm 6$~m\textsuperscript{2}/s for R104, and $D=18\pm 9$~m\textsuperscript{2}/s for R700. 
The measured value of $D$ for R700 is in excellent agreement with previous wavelength-dependent 
measurements performed in transmission~\cite{SperlingPhD2015}: $D \approx 18$~m\textsuperscript{2}/s for $\lambda = 700$~nm~\cite{Schertel2019}. Other previously reported values are $D\sim 20$~m\textsuperscript{2}/s for Ti-Pure and $D\sim 18-38$~m\textsuperscript{2}/s for R104~\cite{SperlingPhD2015}, which differ from ours, but were performed for wavelengths at the lowest range of our experimental spectrum. Moreover, the relative values of $D$ that we obtain are logical in light of the differing values of $k\ell^*$ reported for the three samples~\cite{SperlingPhD2015,Schertel2019}. 

For $t<\tau_c$, $w^2(t)$ appears to be first quasi-ballistic ($t<$ 50 fs), and then subdiffusive~\cite{Hu2008}. The scaling of $P_R(t)$ (Fig.\ \ref{fig:RS}) indicates that diffusion is renormalized in this time range; thus, the behavior of $w^2(t)$ can be predicted by substituting $D(t)$ (Eq.~\ref{locD}) for $D_B$ in Eq.~\ref{eq:w2_4DBt}, giving
\begin{equation}
\label{w2_loc}
    w^2(t)=\frac{4}{\sqrt[3]{6}}(D_0\ell^*t)^{2/3}+w_0^2,
\end{equation}
where $w_0^2$ is the width of the diffuse halo extrapolated to time $t=0$. Fitting the  experimental $w^2(t)$ curves with Eq.~\ref{w2_loc} confirms the scaling of $w^2(t)$ as $t^{2/3}$ for $t<\tau_c$. The subdiffusion-diffusion 
\lc{crossover} can be more clearly seen by plotting $\Delta w^2=w^2-w_0^2$
on a  log-log scale -- this is shown for R700 in the inset of Fig.\ \ref{fig:w2}. 
Fitting data with Eq.~\ref{w2_loc} (using the value of $\ell^*$ measured for R700 at $\lambda=700$ nm~\cite{Schertel2019})  also gives an estimate of $D_0$; remarkably, we find $D_0\sim 15$~m\textsuperscript{2}/s, which agrees within error with $D\sim 18 \pm 9$~m\textsuperscript{2}/s measured from a linear fit of $w^2(t)$ in the diffuse regime (Eq.\ \ref{eq:w2_4DBt}).

\section{Discussion}

\lc{We have observed a clear crossover from a subdiffusive regime at early times to conventional diffusion at later times. By comparing both the return probability and $w^2(t)$ with predictions from the scaling theory of localization, we can conclude that the early-time subdiffusion arises from the renormalization of the diffusion coefficient, and thus falls into the more specific category of renormalized diffusion (a precursor to Anderson localization). This conclusion is supported by the extremely strong scattering that we, and previous groups, have observed in the white powders under investigation, and by the near-field coupling model proposed by Dogariu et al.~\cite{Naraghi2015} to explain the subdiffusion-diffusion crossover.}  

\lc{It is important to differentiate between the subdiffusion-diffusion crossover associated with $\tau_c$, and the continuous phase transition of 3D  Anderson localization. Going through the Anderson transition, the changing correlation/localization length $\xi$ allows the tracking of the transition. While we can observe a crossover in transport behaviour on either side of the subdiffusion-diffusion 
\lc{crossover}, and hence identify the location of the 
\lc{crossover} ($\tau_c$), this crossover can not be characterized any one parameter in the way that a true phase transition can. 
However, a more quantitative study of the crossover time and its scaling with the different parameters of Eq.\ \ref{eq:tau_c} is an interesting perspective of this work. As such a study would require precise control over the near-field cross-section -- a feat which is extremely difficult experimentally -- numerical simulation would be a more appropriate tool for this future work.}

Because $w^2(t)$ does not plateau before $\tau_c$ is reached, we can have reasonable confidence that the regime of strong localization was not reached~\cite{Hu2008,Cobus2018}; we also do not observe any unexpected scaling of $P_R(t)$ which might result from multifractal effects at the mobility edge (diffusion-localization \lc{phase} transition)~\cite{Akridas-Morel2019}.
However, with the scaling-theory-based analysis presented here, we can not know how close to the mobility edge the system came before $\tau_c$.
This last question is of great interest, as it remains unclear whether 3D localization of light can never be reached because the energy leak destroys the requisite long scattering paths, or if this effect can be overcome by structuring disorder, as a way of minimizing $\rho$~\cite{Escalante2017,Skipetrov2019,Andreoli2021}. Interestingly, 3D AL has been recently predicted in hyperuniform dielectric networks~\cite{Haberko2020}, a new class of highly correlated but disordered photonic band gap materials. More generally, correlated disorder implies a complex transport phase diagram~\cite{FroufePerez2017} whose transitions could in principle be revealed by our passive imaging method independently from absorption, fluorescence and noise issues. Indeed, the acquisition of the reflection matrix enables a frequency-resolved study of transport properties in post-processing~\cite{Aubry2014,Cobus2016}. This is a key aspect as it can lead to an estimation of the critical exponents governing such phase transitions~\cite{Ghosh2015} -- a holy grail of the condensed matter and wave physics communities. 

\lc{A remaining question concerns the transport of light at very early times. We do not observe a 
\lc{crossover} from conventional to 
\lc{subdiffusion} at very early times, as predicted by Naraghi et al.~\cite{Naraghi2016}, leading us to believe that it does not exist. The earliest measured point of $w^2(t)$, at $t\sim50$~fs, is close to the ballistic light line ${w(t)}=c_0t$ (Fig.~4 of the accompanying paper). Only super-diffusive -- if not ballistic -- transport could account for such rapid growth of the diffuse halo. One possible explanation is the existence of ballistic waves propagating at the surface of the scattering sample; at early times, this contribution would dominate the observed dynamics of the diffuse halo.}

\section{Conclusions and perspectives}

In summary, we have quantified the spatio-temporal optical energy transport in a strongly scattering regime across a wide range of time scales. We observe a 
\lc{crossover} between a regime of continuously renormalized diffusion at early times, and a conventional diffusion regime at long times. This 
\lc{crossover} was previously predicted to occur due to near-field dipole-dipole coupling between scatterers, which redirects energy from long recurrent scattering paths, destroying AL~\cite{Naraghi2016}. Our measurements offer the best evidence to date for this effect. Nevertheless, these results do not close the debate about AL for light since a structured disorder may cancel the detrimental impact of near-field coupling. In that quest, passive matrix imaging constitutes an excellent platform to provide an unambiguous proof for AL and, more generally, to investigate the fascinating transport properties of light in correlated disordered media~\cite{Vynck2021}.

\acknowledgments
The authors would like to thank Nicolas Lequeux for his help in fabricating the samples, Lukas Schertel for his advice and support, and Sergey Skipetrov and Bart van Tiggelen for fruitful discussions. 
This project has received funding from the Labex WIFI (Laboratory of Excellence within the French Program Investments for the Future, ANR-10-LABX-24 and ANR-10-IDEX-0001-02 PSL*) and the Agence Nationale de la Recherche (ANR-14-CE26-0032, Research Project LOVE). L. A. C. acknowledges financial support from the European Union’s Horizon 2020 research and innovation programme under the Marie Sklodowska-Curie grant agreement No. 744840. A. A. acknowledges financial support from the European Research Council (ERC) under the European Union’s Horizon 2020 research and innovation programme (grant agreement No. 819261).

\appendix

\section{Physical interpretation of our passive imaging method}
\label{sec:Theory0}

{The} scattered wave-field $\psi_c$ measured by the camera 
{is now} investigated in the temporal Fourier domain.  
{Using the Rayleigh-Sommerfeld integral, $\psi_c$ can be expressed as ~\cite{Goodman}}
\begin{equation}
   \psi_c(\mathbf{r_A},\omega)= jk \int_S d\mathbf{r}\ H(\mathbf{r}-\mathbf{r}_A,\omega) \psi_s(\mathbf{r},\omega) ,
\end{equation}
{where $S$ is the surface of the scattering medium, $\omega$ is frequency, $k=\omega/c$ is the optical wave number, $c$ is optical wave-speed, $\psi_s(\mathbf{r},\omega)$ is the optical wave-field at the surface of the sample, and $H(\mathbf{r}-\mathbf{r}_A,\omega)$ is} the spatial impulse response between the sample and the camera. In the following, given the limited bandwidth of the light source ($\Delta \omega/\omega \sim 30\% $), the impulse response $H$ will be taken as independent 
{of} frequency $\omega$. 
The time-derivative of the mutual coherence function of this wave-field 
can be expressed as 
\begin{align}
\label{eq:CAB}
R(\mathbf{r}_A,\mathbf{r}_B,\omega) &\equiv j \omega \langle\psi_c(\mathbf{r_A},\omega) \psi_c^*(\mathbf{r_B},\omega)\rangle_t \nonumber \\ &= j \omega k^2 \int_S d\mathbf{r}_1 \int_S d\mathbf{r_2}\ H(\mathbf{r}_1-\mathbf{r}_A) 
\nonumber \\
&\times H^*(\mathbf{r}_2-\mathbf{r}_B)  \langle\psi_s(\mathbf{r}_1,\omega) \psi_s^*(\mathbf{r_2},\omega)\rangle  ,
\end{align}
where the symbol $\langle \cdots \rangle$ denotes an ensemble average. 
For an ambient wave field $\psi_s(\mathbf{r},\omega)$ equipartitioned in energy in 
{phase space, the} fluctuation-dissipation theorem implies that the time-derivative of the mutual coherence function $\langle\psi_s(\mathbf{r}_1,\omega) \psi_s^*(\mathbf{r}_2,\omega)\rangle$ converges towards the imaginary part of the Green's function between $\mathbf{r}_1$ and $\mathbf{r}_2$~\cite{Weaver2004}: 
\begin{align}
\label{eq:C_RR}
    j \omega \langle\psi_s(\mathbf{r}_1,\omega) \psi_s^*(\mathbf{r}_2,\omega)\rangle &= \text{Im}G(\mathbf{r}_1,\mathbf{r}_2,\omega) 
    \nonumber \\
&=[G(\mathbf{r}_1,\mathbf{r}_2,\omega)-G^*(\mathbf{r}_1,\mathbf{r}_2,\omega)] .
\end{align} 
{Here,} $G(\mathbf{r}_1,\mathbf{r}_2,\omega)$ and $G^*(\mathbf{r}_1,\mathbf{r}_2,\omega)$ stand for the causal (retarded) and anti-causal(advanced) parts of the Green's function, respectively.  In our measurement, only the causal part ($t>0$) of the correlation signal $C_T(\mathbf{r}_A,\mathbf{r}_B,t>0)$ is recorded. Hence, $\text{Im}G(\mathbf{r}_1,\mathbf{r}_2,\omega)$ can be replaced by the retarded Green's function in Eq.~\ref{eq:C_RR}, such that
\begin{align}
\label{eq:CAB_ImG}
  R(\mathbf{r_A},\mathbf{r_B},\omega) &= k^2  \int_S d\mathbf{r}_1 \int_S d\mathbf{r}_2 H(\mathbf{r}_1-\mathbf{r}_A)  \nonumber \\
&\times G(\mathbf{r}_1,\mathbf{r}_2,\omega)  H^*(\mathbf{r}_2-\mathbf{r}_B).
\end{align}
This equation can be given the following physical interpretation by reading the integrands from left to right{:} $H(\mathbf{r}_1-\mathbf{r_A}) $ 
{describes} the amplitude distribution at point $\mathbf{r}_1$ of an incident wave-field generated by a virtual source located at $\mathbf{r}_A$, 
$G(\mathbf{r}_1,\mathbf{r}_2,\omega)$ describes wave propagation in the sample
from $\mathbf{r}_1$ to $\mathbf{r}_2$ where the first and last scattering events occur, and 
$H^*(\mathbf{r}_2-\mathbf{r_B})$ describes the
propagation between the last scattering event and the virtual detector at $\mathbf{r_B}$. The covariance matrix $\mathbf{R}(\omega)=[ R(\mathbf{r_A},\mathbf{r_B},\omega) ]$ can thus be seen as the reflection matrix of the scattering medium measured in real space.\\

\section{Theoretical description of the mean intensity profile}
\label{sec:TheoryI}

In this work, we are interested in the spatio-temporal evolution of the mean backscattered intensity. 
{A theoretical prediction for this quantity can be derived by} considering the ensemble averaged intensity of the time-dependent mutual coherence function:
\begin{equation}
\label{eq:Ikl}
    I(\mathbf{r}_A,\mathbf{r}_B,t)= \langle |R(\mathbf{r_A},\mathbf{r_B},t) |^2\rangle .
\end{equation}
To express $I(\mathbf{r}_A,\mathbf{r}_B,t)$, we first consider its temporal frequency counterpart, $I(\mathbf{r}_A,\mathbf{r}_B, \Omega)$:
\begin{equation}
    I(\mathbf{r}_A,\mathbf{r}_B, \Omega)=\int dt I(\mathbf{r}_A,\mathbf{r}_B,t) e^{-i\Omega t}.
\end{equation}
Using Eq.~\ref{eq:Ikl}, $ I(\mathbf{r}_A,\mathbf{r}_B, \Omega)$ can be rewritten as
\begin{equation}
\label{C0}
    I(\mathbf{r}_A,\mathbf{r}_B,  \Omega)=  \langle R(\mathbf{r}_A,\mathbf{r}_B,  \omega) R^*(\mathbf{r}_A,\mathbf{r}_B,  \omega- \Omega)  \rangle_{\omega}  .
\end{equation}
Injecting Eq.~\ref{eq:CAB_ImG} into 
{Eq.\ \ref{C0} gives} 
\begin{widetext}
\begin{align}
\label{eq:CAB_int}
    I(\mathbf{r}_A,\mathbf{r}_B,  \Omega)= &  k^4 \int_S d\mathbf{r}_1 \int_S d\mathbf{r}_2 \int_S d\mathbf{r}^\prime_1 \int_S d\mathbf{r}^\prime_2   {H}(\mathbf{r}_1-\mathbf{r}_A){H}^*(\mathbf{r}^\prime_1-\mathbf{r}_A)  \nonumber \\
& \times \left \langle {G}(\mathbf{r}_1,\mathbf{r}_2,\omega)
{G}^*(\mathbf{r}^\prime_1,\mathbf{r}^\prime_2,\omega-\Omega) \right \rangle_\omega  {H}^*(\mathbf{r}_2-\mathbf{r}_B){H}(\mathbf{r}^\prime_2-\mathbf{r}_B) .
\end{align}
\end{widetext}

In the weak scattering regime ($k\ell^* >>1$), most contributions to the correlation function of the Green's function at the surface of the scattering medium, $\left \langle G(\mathbf{r}_1 ,\mathbf{r}_2,\omega ) G^*(\mathbf{r}'_1,\mathbf{r}'_2,\omega -\Omega ) \right \rangle_\omega$, will cancel out in the above ensemble average. The only contributions to survive this average are those for which the wave and its complex conjugate experience identical paths. This condition is achieved if the wave and the complex conjugate visit the same scatterers either in the same order (ladder diagrams), or in reversed order (maximally crossed diagrams). The correlation function can thus be decomposed into two terms~\cite{Akkermans2007}:
\begin{align}
    \label{corr_real}
    & \langle G(\mathbf{r}_1  ,\mathbf{r}_2, \omega ) G^*(\mathbf{r}'_1  ,\mathbf{r}'_2,\omega -\Omega ) \rangle_\omega  
    \nonumber \\
    & =   \frac{c}{k^4}P(\mathbf{r}_1  ,\mathbf{r}_2,\Omega )  {[}      \delta(\mathbf{r}_1-\mathbf{r}^\prime_1)\delta(\mathbf{r}_2-\mathbf{r}^\prime_2)
    \nonumber \\
    & + \delta(\mathbf{r}_1-\mathbf{r}^\prime_2)\delta(\mathbf{r}^\prime_1-\mathbf{r}_2) {]},
\end{align}
where $P(\mathbf{r}_1,\mathbf{r}_2,\Omega)$ is an energy density. Physically, $P(\mathbf{r}_1,\mathbf{r}_2,\Omega)$ is the Fourier transform of $P(\mathbf{r}_1,\mathbf{r}_2,t)$ -- the probability to find a pulse at point $\mathbf{r}$ and time $t$, after emission of a short pulse at point $\mathbf{r}^\prime$. The first term {of Eq.\ \ref{corr_real}} describes the self-interference of the wave associated with each possible scattering path between $\mathbf{r}_1$ and $\mathbf{r}_2$. The second describes the constructive interference between reciprocal scattering paths between the same points.

Injecting Eq.~\ref{corr_real} into Eq.~\ref{eq:CAB_int} leads to a decomposition of the mean back-scattered {intensity} as the sum of an incoherent ($I_\textrm{inc}$) and a coherent ($I_\textrm{coh}$) component. In the temporal regime, this can be expressed as
\begin{equation}
\label{eq:Iab}
    I(\mathbf{r}_A,\mathbf{r}_B,t) =  I_\textrm{inc}(\mathbf{r}_A,\mathbf{r}_B,t) +I_\textrm{coh}(\mathbf{r}_A,\mathbf{r}_B,t) .
\end{equation}
The incoherent intensity $I_\textrm{inc}$ accounts for the self interference of waves propagating along the same scattering paths, 
\begin{align}
\label{Iinc}
 I_\textrm{inc} ( \mathbf{r}_A,\mathbf{r}_B,t)  = c   \int_S d\mathbf{r}_1 \int_S d\mathbf{r}_2 \ |{H}(\mathbf{r}_1-\mathbf{r}_A)|^2 
 \nonumber \\
 \times {P}(\mathbf{r}_2,\mathbf{r}_1,t)
|{H}(\mathbf{r}_2-\mathbf{r}_B)|^2 ,  
\end{align}
{while} the coherent intensity $I_\textrm{coh}$ 
{is associated with} the interference of waves 
following reciprocal scattering paths,
\begin{align}
\label{Icoh}
 {I_\textrm{coh} (\mathbf{r}_A,\mathbf{r}_B,t)}
 \nonumber \\
 &=  c  \int_S d\mathbf{r}_1 \int_S d\mathbf{r}_2  {H}(\mathbf{r}_1-\mathbf{r}_A)
 {H}(\mathbf{r}_1-\mathbf{r}_B) 
 \nonumber \\
 &\times {P}(\mathbf{r}_2,\mathbf{r}_1,t)
{H}^*(\mathbf{r}_2-\mathbf{r}_A) {H}^*(\mathbf{r}_2-\mathbf{r}_B).
\end{align} 
This term accounts for the so-called coherent backscattering phenomenon. 
{To simplify the preceeding expressions, the medium can be} 
assumed to be statistically homogeneous such that $P$ is invariant by translation: $P(\mathbf{r}_2,\mathbf{r}_1,t)=P(\mathbf{r}_2-\mathbf{r}_1,t)$. 
{Then, the} 
incoherent intensity (Eq.~\ref{Iinc}) can be simplified 
{to}:
\begin{equation}
\label{Iinc2}
 I_\textrm{inc} ( \Delta r,t) = c  \left [|H|^2 \stackrel{\Delta \mathbf{r}}{\circledast} |H|^2 \stackrel{\Delta \mathbf{r}}{\circledast} P(\Delta \mathbf{r},t)\right ],
\end{equation}
where $\Delta \mathbf{r}=\mathbf{r}_B-\mathbf{r}_A$ is the relative position between the virtual source and detector and the symbol $\stackrel{\Delta \mathbf{r}}{\circledast}$ stands for the correlation product over $\Delta \mathbf{r}$. 

In the weak disorder regime {($k\ell^* >>1$)}, $P(\Delta \mathbf{r},t)$ is the solution to the diffusion equation, 
in which diffusivity {$\mathcal{D}$} is constant and corresponds to the Boltzmann diffusion coefficient $\mathcal{D}=D_B=v_E\ell^*/3$, where $v_E$ is the energy transport velocity. In this regime~\cite{Patterson1989}, 
\begin{equation}
\label{time_P}
P(\Delta \mathbf{r},t)=\frac{1}{\pi^{3/2} w^{3}(t)}\exp{\left[ -\frac{ |\Delta \mathbf{r}|^2}{w^2(t)} \right]}
\end{equation}

\begin{figure*}
	\centering\includegraphics[width=0.85\textwidth]{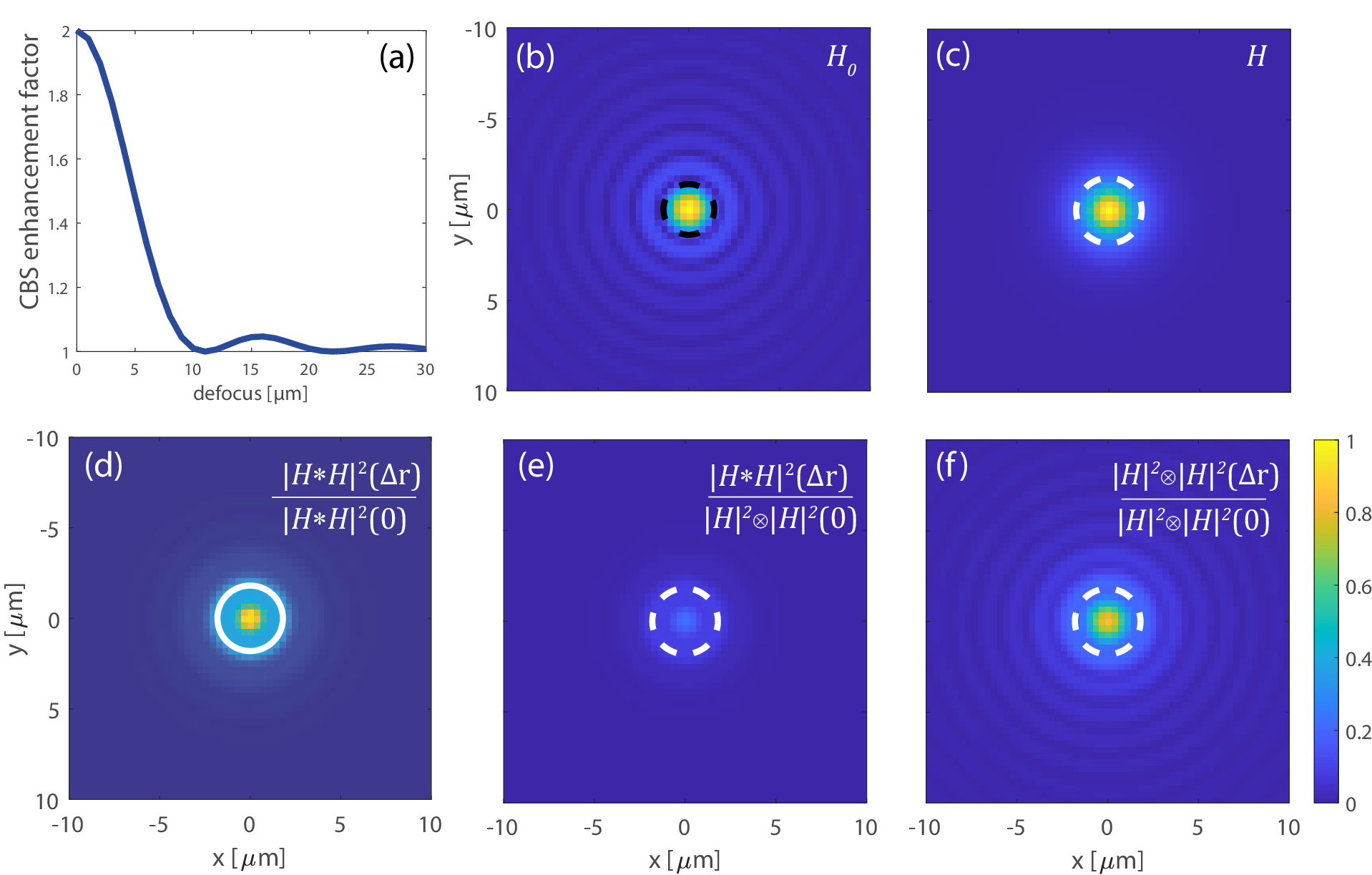}
	\caption{(a) Evolution of the CBS enhancement factor $A$ as a function of defocus. (b) Diffraction-limited PSF $H_0(\mathbf{r})$. The dashed black circle accounts for the diffraction limited focal spot of radius $\delta r \sim \lambda/(2NA) \sim 1.4$ $\mu$m. (c) PSF $H(\mathbf{r})$ for a defocus $z=7$ $\mu$m.  (d,e) Corresponding {CBS} peak $|H \stackrel{\Delta \mathbf{r}}{\ast} H|^2(\Delta \mathbf{r})$ normalized by its maximum (c) and the incoherent PSF  $|H|^2 \stackrel{\Delta \mathbf{r}}{\circledast} |H|^2$ at $\Delta r$=0. (f) Corresponding incoherent PSF $|H|^2 \stackrel{\Delta \mathbf{r}}{\circledast} |H|^2(\Delta \mathbf{r})$. In panels (c)-(e), the white dashed circle accounts for the defocused focal spot of radius $\delta r \sim z NA/\sqrt{1-NA^2} \sim$ 1.8 $\mu$m. }
	\label{fig:defocus}
\end{figure*}

As soon as $w^2(t)$ is much larger than the spatial extent $\delta r^2$ of $|H|^2$ ($w^2(t)>> \delta r^2$), the incoherent intensity (Eq.~\ref{Iinc2}) is a reliable estimator of $P(\Delta \mathbf{r},t)$ :
\begin{equation}
\label{Iinc3}
 I_\textrm{inc} ( \Delta \mathbf{r},t) \underset{\delta r^2 << w^2(t)}{\sim} c  P(\Delta \mathbf{r},t).
\end{equation}
Under the same condition {,} 
${P}(\mathbf{r}_2-\mathbf{r}_1,t)$ can be replaced by ${P}(\mathbf{0},t)$ in the integrand of Eq.~\ref{Icoh}. The expression of $ I_\textrm{coh}$ then simplifies into:
\begin{equation}
\label{Icoh2}
 I_\textrm{coh} ( \Delta \mathbf{r},t) \underset{\delta r^2 << w^2(t)}{\sim} c  P( \mathbf{0},t) \times |H \stackrel{\Delta \mathbf{r}}{\ast} H|^2 (\Delta \mathbf{r}) ,
\end{equation}
where the symbol $\stackrel{\Delta \mathbf{r}}{\ast}$ stands for a convolution product over $\Delta \mathbf{r}$.
In our experimental configuration where {$\delta r^2 = {3.2}$~$\mu\text{m}^2$, this condition is already reached at the earliest measured times of flight; thus,} the shape of the {CBS} peak is 
governed by the coherent PSF $ |H \stackrel{\Delta \mathbf{r}}{\ast} H|^2 (\Delta \mathbf{r})$ {for all times}.

Using Eqs.~\ref{Iinc2} and \ref{Icoh2}, a theoretical expression for the coherent backscattering enhancement $A$ can be derived:
\begin{align}
    A&=1+\frac{I_\textrm{coh}(\Delta r=0)}{I_\textrm{inc}(\Delta r=0)}
    \nonumber \\&=1+\frac{|H \stackrel{\Delta \mathbf{r}}{\ast} H|^2  (\Delta {r}=0)}{|H|^2 \stackrel{\Delta \mathbf{r}}{\circledast} |H|^2 (\Delta {r}=0)}.
\end{align}
The shape of the {CBS} peak is given by the function
\begin{equation}
    F(\Delta r)=\frac{|H \stackrel{\Delta \mathbf{r}}{\ast} H|^2  (\Delta {r})}{|H \stackrel{\Delta \mathbf{r}}{\ast} H|^2  (\Delta {r}=0)}.
\end{equation}
In the absence of aberrations, the PSF is only limited by diffraction: $H\equiv H_0$, with $H_0=\sqrt{2}J_1(k NA \Delta r)/(k NA \Delta r)$ (Fig.~\ref{fig:defocus}(b)). In this ideal case, the CBS enhancement $A$ is equal to 2 since $|H_0|^2 \stackrel{\Delta \mathbf{r}}{\circledast} |H_0|^2 (\Delta {r}=0)\equiv |H_0 \stackrel{\Delta \mathbf{r}}{\ast} H_0|^2  (\Delta {r}=0)$. The CBS peak then coincides with the Airy disk: $F(\Delta r)=|H_0(\Delta r)|^2$.

In the real world, any imaging system suffers from aberrations. {Relying on a simple Fourier optics model~\cite{Barolle2021}, Fig.~\ref{fig:defocus}(a)} shows, for instance, the effect of a defocus on the {CBS} enhancement. {The enhancement factor} $A$ falls off rapidly with the defocus distance $d$ and cancels for $d=n z_R/2$, {where $n$ is} a positive integer and $z_R=2\lambda/NA^2\sim 22$~$\mu$m {is} the Rayleigh range or depth-of-field. The weak {CBS} enhancement ($A\sim 0.1-0.2)$ found in our experiments would correspond to a defocus of $d\sim 7$~$\mu$m. The corresponding PSF $H$, the associated {CBS} peak $F(\Delta \mathbf{r})$ and the incoherent PSF $|H|^2 \stackrel{\Delta \mathbf{r}}{\circledast} |H|^2(\Delta \mathbf{r})$ are displayed for this value of defocus in Fig.~\ref{fig:defocus}(c), (d,e) and (f), respectively. This figure illustrates the drastic effect of defocus on the {CBS} peak with respect to the incoherent PSF. The spatial extent $\delta r$ of each quantity is roughly equal to the transverse resolution in presence of a defocus $z$: $\delta r\sim z NA/\sqrt{1-NA^2} \sim 1.8$ $\mu$m~\cite{Barolle2021}. Higher-order aberrations such as astigmatism could also contribute to the weak value of CBS enhancement observed in our experiments.

\section{Fitting normalized intensity profiles with theory}
\label{sec:fittingIwiththeories}

\begin{figure*}
\centering
	\includegraphics[width=\textwidth]{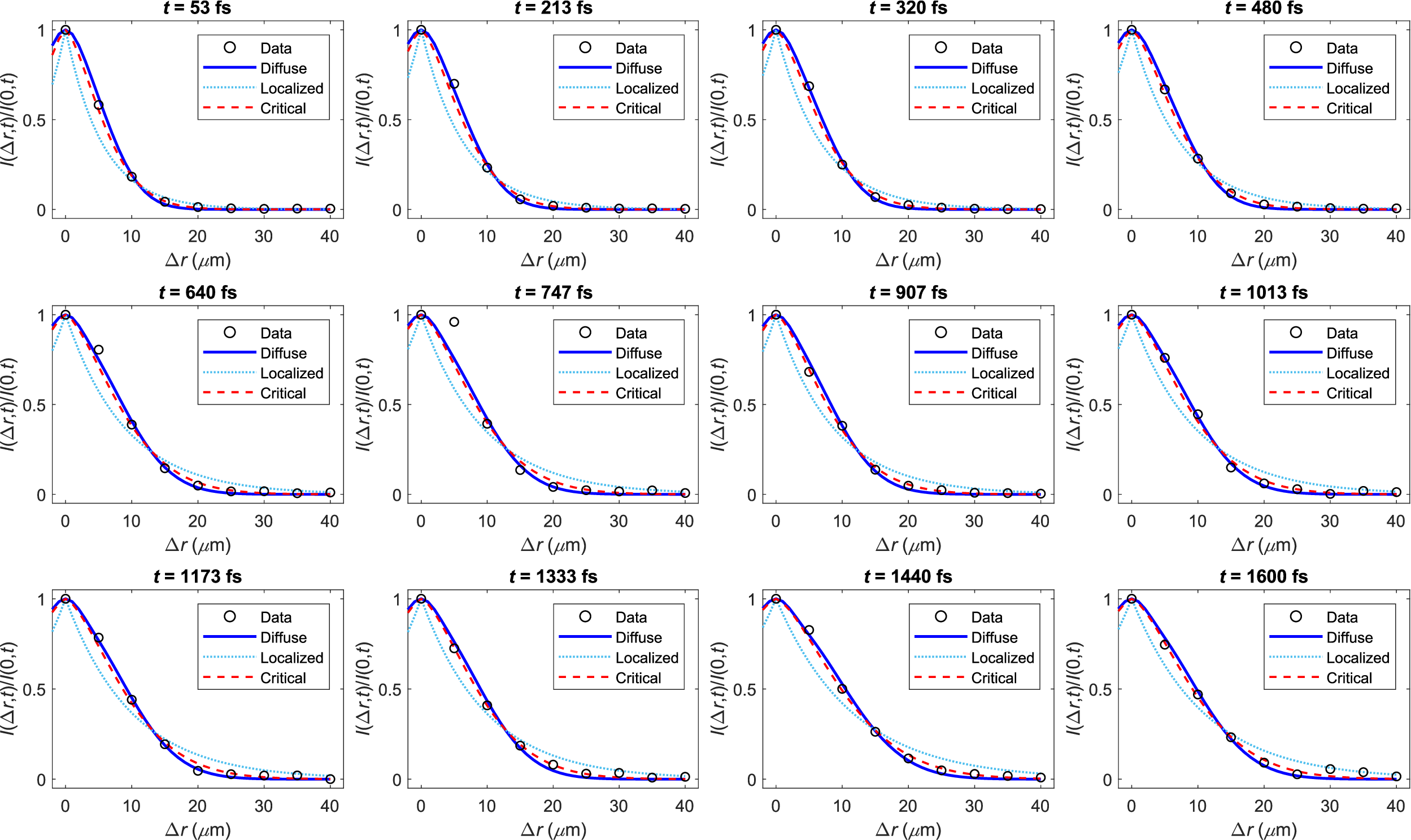}
	\caption{Representative  normalized intensity profiles are shown for sample R700. Of the 30 points in time measured, 12 are shown here which span the entire measurement range. For each time, data were fit with theoretical predictions for the diffuse (Eq.\ \ref{eq:intensity_profile_gauss}, blue solid lines), critical (Eq.\ \ref{eq:intensity_profile_crit}, red dashed lines) and localized (Eq.\ \ref{eq:intensity_profile_gauss}, cyan dotted lines) regimes. }
	\label{fig:fits_diff_loc_crit}
\end{figure*}

For conventional diffusion, the normalized intensity profiles $I(\Delta r,t)/I(0,t)$ can be described by Eq.\ 3, reproduced here:
\begin{align}
\label{eq:intensity_profile_gauss}
   \frac{ I(\Delta r,t)}{I(0,t)} = \frac{1}{A}e^{-\Delta r^2/w^2(t)}+\left(1-\frac{1}{A}\right)F(\Delta r).   
\end{align}
For renormalized diffusion, however, the spatial shape of the profiles may deviate from a Gaussian. In the limit of strong localization in an infinite medium, this shape is exponential:
\begin{align}
\label{eq:intensity_profile_loc}
   \frac{ I(\Delta r,t)}{I(0,t)} \sim \frac{1}{A}e^{-\Delta r/\xi}+\left(1-\frac{1}{A}\right)F(\Delta r),   
\end{align}
where $\xi$ is the localization length defining the average spatial confinement of the wave energy. At the mobility edge, we might expect~\cite{Lemarie2010}
\begin{align}
\label{eq:intensity_profile_crit}
   \frac{ I(\Delta r,t)}{I(0,t)} \sim \frac{1}{A}e^{-\al{\alpha}\Delta r^{3/2}/t^{1/2}}+\left(1-\frac{1}{A}\right)F(\Delta r).   
\end{align}
\al{with $\alpha$ a constant to be determined.} Fitting our experimental data with these three expressions, we find poor agreement for the localized limit (Eq.\ \ref{eq:intensity_profile_loc}) at all times $t$ \lc{(note that, for all fits, the enhancement coefficient $A$ was a free parameter, allowed to vary from $1$ to $3$.)}. As shown in Fig.\ \ref{fig:fits_diff_loc_crit}, both diffuse and critical regime expressions fit the experimental data well at all times. The sparsity of the data points in space means that neither model fits significantly better than the other. 
For the purposes of characterizing the spatio-temporal spread of $I_{inc}$, it is therefore valid to use the Gaussian model given by Eq.\ \ref{eq:intensity_profile_gauss} to estimate $w^2(t)$. In future work, better characterization of the shape of $I(\Delta r,t)$ may help to identify if, in a certain time range, the system is near enough to the mobility edge for a scaling like Eq.\ \ref{eq:intensity_profile_crit} to apply. 


\begin{thebibliography}{70}%
\makeatletter
\providecommand \@ifxundefined [1]{%
 \@ifx{#1\undefined}
}%
\providecommand \@ifnum [1]{%
 \ifnum #1\expandafter \@firstoftwo
 \else \expandafter \@secondoftwo
 \fi
}%
\providecommand \@ifx [1]{%
 \ifx #1\expandafter \@firstoftwo
 \else \expandafter \@secondoftwo
 \fi
}%
\providecommand \natexlab [1]{#1}%
\providecommand \enquote  [1]{``#1''}%
\providecommand \bibnamefont  [1]{#1}%
\providecommand \bibfnamefont [1]{#1}%
\providecommand \citenamefont [1]{#1}%
\providecommand \href@noop [0]{\@secondoftwo}%
\providecommand \href [0]{\begingroup \@sanitize@url \@href}%
\providecommand \@href[1]{\@@startlink{#1}\@@href}%
\providecommand \@@href[1]{\endgroup#1\@@endlink}%
\providecommand \@sanitize@url [0]{\catcode `\\12\catcode `\$12\catcode
  `\&12\catcode `\#12\catcode `\^12\catcode `\_12\catcode `\%12\relax}%
\providecommand \@@startlink[1]{}%
\providecommand \@@endlink[0]{}%
\providecommand \url  [0]{\begingroup\@sanitize@url \@url }%
\providecommand \@url [1]{\endgroup\@href {#1}{\urlprefix }}%
\providecommand \urlprefix  [0]{URL }%
\providecommand \Eprint [0]{\href }%
\providecommand \doibase [0]{https://doi.org/}%
\providecommand \selectlanguage [0]{\@gobble}%
\providecommand \bibinfo  [0]{\@secondoftwo}%
\providecommand \bibfield  [0]{\@secondoftwo}%
\providecommand \translation [1]{[#1]}%
\providecommand \BibitemOpen [0]{}%
\providecommand \bibitemStop [0]{}%
\providecommand \bibitemNoStop [0]{.\EOS\space}%
\providecommand \EOS [0]{\spacefactor3000\relax}%
\providecommand \BibitemShut  [1]{\csname bibitem#1\endcsname}%
\let\auto@bib@innerbib\@empty
\bibitem [{\citenamefont {Anderson}(1958)}]{Anderson1958}%
  \BibitemOpen
  \bibfield  {author} {\bibinfo {author} {\bibfnamefont {P.~W.}\ \bibnamefont
  {Anderson}},\ }\href@noop {} {\bibfield  {journal} {\bibinfo  {journal}
  {Phys. Rev.}\ }\textbf {\bibinfo {volume} {109}},\ \bibinfo {pages} {1492}
  (\bibinfo {year} {1958})}\BibitemShut {NoStop}%
\bibitem [{\citenamefont {John}(1984)}]{John1984}%
  \BibitemOpen
  \bibfield  {author} {\bibinfo {author} {\bibfnamefont {S.}~\bibnamefont
  {John}},\ }\href@noop {} {\bibfield  {journal} {\bibinfo  {journal} {Phys.
  Rev. Lett.}\ }\textbf {\bibinfo {volume} {53}},\ \bibinfo {pages} {2169}
  (\bibinfo {year} {1984})}\BibitemShut {NoStop}%
\bibitem [{\citenamefont {John}(1991)}]{John1991}%
  \BibitemOpen
  \bibfield  {author} {\bibinfo {author} {\bibfnamefont {S.}~\bibnamefont
  {John}},\ }\href@noop {} {\bibfield  {journal} {\bibinfo  {journal} {Phys.
  Today}\ }\textbf {\bibinfo {volume} {44}},\ \bibinfo {pages} {32} (\bibinfo
  {year} {1991})}\BibitemShut {NoStop}%
\bibitem [{\citenamefont {Evers}\ and\ \citenamefont
  {Mirlin}(2008)}]{Evers2008}%
  \BibitemOpen
  \bibfield  {author} {\bibinfo {author} {\bibfnamefont {F.}~\bibnamefont
  {Evers}}\ and\ \bibinfo {author} {\bibfnamefont {A.~D.}\ \bibnamefont
  {Mirlin}},\ }\href@noop {} {\bibfield  {journal} {\bibinfo  {journal} {Rev.
  Mod. Phys.}\ }\textbf {\bibinfo {volume} {80}},\ \bibinfo {pages} {1355}
  (\bibinfo {year} {2008})}\BibitemShut {NoStop}%
\bibitem [{\citenamefont {Rosenbaum}\ \emph {et~al.}(1980)\citenamefont
  {Rosenbaum}, \citenamefont {Andres}, \citenamefont {Thomas},\ and\
  \citenamefont {Bhatt}}]{Rosenbaum1980}%
  \BibitemOpen
  \bibfield  {author} {\bibinfo {author} {\bibfnamefont {T.~F.}\ \bibnamefont
  {Rosenbaum}}, \bibinfo {author} {\bibfnamefont {K.}~\bibnamefont {Andres}},
  \bibinfo {author} {\bibfnamefont {G.~A.}\ \bibnamefont {Thomas}},\ and\
  \bibinfo {author} {\bibfnamefont {R.~N.}\ \bibnamefont {Bhatt}},\ }\href@noop
  {} {\bibfield  {journal} {\bibinfo  {journal} {Phys. Rev. Lett.}\ }\textbf
  {\bibinfo {volume} {45}},\ \bibinfo {pages} {1723} (\bibinfo {year}
  {1980})}\BibitemShut {NoStop}%
\bibitem [{\citenamefont {Lee}\ and\ \citenamefont
  {Ramakrishnan}(1985)}]{Lee1985}%
  \BibitemOpen
  \bibfield  {author} {\bibinfo {author} {\bibfnamefont {P.~A.}\ \bibnamefont
  {Lee}}\ and\ \bibinfo {author} {\bibfnamefont {T.~V.}\ \bibnamefont
  {Ramakrishnan}},\ }\href@noop {} {\bibfield  {journal} {\bibinfo  {journal}
  {Rev. Mod. Phys.}\ }\textbf {\bibinfo {volume} {57}},\ \bibinfo {pages} {287}
  (\bibinfo {year} {1985})}\BibitemShut {NoStop}%
\bibitem [{\citenamefont {Wiersma}\ \emph {et~al.}(1997)\citenamefont
  {Wiersma}, \citenamefont {Bartolini}, \citenamefont {Lagendijk},\ and\
  \citenamefont {Righini}}]{Wiersma1997}%
  \BibitemOpen
  \bibfield  {author} {\bibinfo {author} {\bibfnamefont {D.~S.}\ \bibnamefont
  {Wiersma}}, \bibinfo {author} {\bibfnamefont {P.}~\bibnamefont {Bartolini}},
  \bibinfo {author} {\bibfnamefont {A.}~\bibnamefont {Lagendijk}},\ and\
  \bibinfo {author} {\bibfnamefont {R.}~\bibnamefont {Righini}},\ }\href@noop
  {} {\bibfield  {journal} {\bibinfo  {journal} {Nature}\ }\textbf {\bibinfo
  {volume} {390}},\ \bibinfo {pages} {671 } (\bibinfo {year}
  {1997})}\BibitemShut {NoStop}%
\bibitem [{\citenamefont {Aegerter}\ \emph {et~al.}(2006)\citenamefont
  {Aegerter}, \citenamefont {St{\"{o}}rzer},\ and\ \citenamefont
  {Maret}}]{Aegerter2006}%
  \BibitemOpen
  \bibfield  {author} {\bibinfo {author} {\bibfnamefont {C.~M.}\ \bibnamefont
  {Aegerter}}, \bibinfo {author} {\bibfnamefont {M.}~\bibnamefont
  {St{\"{o}}rzer}},\ and\ \bibinfo {author} {\bibfnamefont {G.}~\bibnamefont
  {Maret}},\ }\href {https://doi.org/10.1209/epl/i2006-10144-3} {\bibfield
  {journal} {\bibinfo  {journal} {Europhys. Lett.}\ }\textbf
  {\bibinfo {volume} {75}},\ \bibinfo {pages} {562} (\bibinfo {year}
  {2006})}\BibitemShut {NoStop}%
\bibitem [{\citenamefont {St\"orzer}(2006)}]{StorzerPhD2006}%
  \BibitemOpen
  \bibfield  {author} {\bibinfo {author} {\bibfnamefont {M.}~\bibnamefont
  {St\"orzer}},\ } {\bibinfo {title} {{Anderson Localization of Light}}},\
  \href@noop {} {Ph.D. thesis},\ \bibinfo  {school} {Universit\"at Konstanz}
  (\bibinfo {year} {2006})\BibitemShut {NoStop}%
\bibitem [{\citenamefont {Sperling}\ \emph {et~al.}(2013)\citenamefont
  {Sperling}, \citenamefont {B\"uhrer}, \citenamefont {Aegerter},\ and\
  \citenamefont {Maret}}]{Sperling2013}%
  \BibitemOpen
  \bibfield  {author} {\bibinfo {author} {\bibfnamefont {T.}~\bibnamefont
  {Sperling}}, \bibinfo {author} {\bibfnamefont {W.}~\bibnamefont {B\"uhrer}},
  \bibinfo {author} {\bibfnamefont {C.~M.}\ \bibnamefont {Aegerter}},\ and\
  \bibinfo {author} {\bibfnamefont {G.}~\bibnamefont {Maret}},\ }\href@noop {}
  {\bibfield  {journal} {\bibinfo  {journal} {Nat. Photonics}\ }\textbf
  {\bibinfo {volume} {7}},\ \bibinfo {pages} {48} (\bibinfo {year}
  {2013})}\BibitemShut {NoStop}%
\bibitem [{\citenamefont {Scheffold}\ \emph {et~al.}(1999)\citenamefont
  {Scheffold}, \citenamefont {Lenke}, \citenamefont {Tweer},\ and\
  \citenamefont {Maret}}]{Scheffold1999}%
  \BibitemOpen
  \bibfield  {author} {\bibinfo {author} {\bibfnamefont {F.}~\bibnamefont
  {Scheffold}}, \bibinfo {author} {\bibfnamefont {R.}~\bibnamefont {Lenke}},
  \bibinfo {author} {\bibfnamefont {R.}~\bibnamefont {Tweer}},\ and\ \bibinfo
  {author} {\bibfnamefont {G.}~\bibnamefont {Maret}},\ }\href@noop {}
  {\bibfield  {journal} {\bibinfo  {journal} {Nature}\ }\textbf {\bibinfo
  {volume} {398}},\ \bibinfo {pages} {206 } (\bibinfo {year}
  {1999})}\BibitemShut {NoStop}%
\bibitem [{\citenamefont {{van der Beek}}\ \emph {et~al.}(2012)\citenamefont
  {{van der Beek}}, \citenamefont {Barthelemy}, \citenamefont {Johnson},
  \citenamefont {Wiersma},\ and\ \citenamefont {Lagendijk}}]{VanDerBeek2012}%
  \BibitemOpen
  \bibfield  {author} {\bibinfo {author} {\bibfnamefont {T.}~\bibnamefont {{van
  der Beek}}}, \bibinfo {author} {\bibfnamefont {P.}~\bibnamefont
  {Barthelemy}}, \bibinfo {author} {\bibfnamefont {P.~M.}\ \bibnamefont
  {Johnson}}, \bibinfo {author} {\bibfnamefont {D.~S.}\ \bibnamefont
  {Wiersma}},\ and\ \bibinfo {author} {\bibfnamefont {A.}~\bibnamefont
  {Lagendijk}},\ }\href@noop {} {\bibfield  {journal} {\bibinfo  {journal}
  {Phys. Rev. B.}\ }\textbf {\bibinfo {volume} {85}},\ \bibinfo {pages}
  {115401} (\bibinfo {year} {2012})}\BibitemShut {NoStop}%
\bibitem [{\citenamefont {Sperling}\ \emph {et~al.}(2016)\citenamefont
  {Sperling}, \citenamefont {Schertel}, \citenamefont {Ackermann},
  \citenamefont {Aubry}, \citenamefont {Aegerter},\ and\ \citenamefont
  {Maret}}]{Sperling2016}%
  \BibitemOpen
  \bibfield  {author} {\bibinfo {author} {\bibfnamefont {T.}~\bibnamefont
  {Sperling}}, \bibinfo {author} {\bibfnamefont {L.}~\bibnamefont {Schertel}},
  \bibinfo {author} {\bibfnamefont {M.}~\bibnamefont {Ackermann}}, \bibinfo
  {author} {\bibfnamefont {G.~J.}\ \bibnamefont {Aubry}}, \bibinfo {author}
  {\bibfnamefont {C.~M.}\ \bibnamefont {Aegerter}},\ and\ \bibinfo {author}
  {\bibfnamefont {G.}~\bibnamefont {Maret}},\ }\href@noop {} {\bibfield
  {journal} {\bibinfo  {journal} {New J. Phys.}\ }\textbf {\bibinfo {volume}
  {18}},\ \bibinfo {pages} {13039} (\bibinfo {year} {2016})}\BibitemShut
  {NoStop}%
\bibitem [{\citenamefont {Hu}\ \emph {et~al.}(2008)\citenamefont {Hu},
  \citenamefont {Strybulevych}, \citenamefont {Page}, \citenamefont
  {Skipetrov},\ and\ \citenamefont {van Tiggelen}}]{Hu2008}%
  \BibitemOpen
  \bibfield  {author} {\bibinfo {author} {\bibfnamefont {H.}~\bibnamefont
  {Hu}}, \bibinfo {author} {\bibfnamefont {A.}~\bibnamefont {Strybulevych}},
  \bibinfo {author} {\bibfnamefont {J.~H.}\ \bibnamefont {Page}}, \bibinfo
  {author} {\bibfnamefont {S.~E.}\ \bibnamefont {Skipetrov}},\ and\ \bibinfo
  {author} {\bibfnamefont {B.~A.}\ \bibnamefont {van Tiggelen}},\ }\href@noop
  {} {\bibfield  {journal} {\bibinfo  {journal} {Nat. Phys.}\ }\textbf
  {\bibinfo {volume} {4}},\ \bibinfo {pages} {945} (\bibinfo {year}
  {2008})}\BibitemShut {NoStop}%
\bibitem [{\citenamefont {Cobus}\ \emph {et~al.}(2016)\citenamefont {Cobus},
  \citenamefont {Skipetrov}, \citenamefont {Aubry}, \citenamefont {van
  Tiggelen}, \citenamefont {Derode},\ and\ \citenamefont {Page}}]{Cobus2016}%
  \BibitemOpen
  \bibfield  {author} {\bibinfo {author} {\bibfnamefont {L.~A.}\ \bibnamefont
  {Cobus}}, \bibinfo {author} {\bibfnamefont {S.~E.}\ \bibnamefont
  {Skipetrov}}, \bibinfo {author} {\bibfnamefont {A.}~\bibnamefont {Aubry}},
  \bibinfo {author} {\bibfnamefont {B.~A.}\ \bibnamefont {van Tiggelen}},
  \bibinfo {author} {\bibfnamefont {A.}~\bibnamefont {Derode}},\ and\ \bibinfo
  {author} {\bibfnamefont {J.~H.}\ \bibnamefont {Page}},\ }\href@noop {}
  {\bibfield  {journal} {\bibinfo  {journal} {Phys. Rev. Lett.}\ }\textbf
  {\bibinfo {volume} {116}},\ \bibinfo {pages} {193901} (\bibinfo {year}
  {2016})}\BibitemShut {NoStop}%
\bibitem [{\citenamefont {Chab{\'{e}}}\ \emph {et~al.}(2008)\citenamefont
  {Chab{\'{e}}}, \citenamefont {Lemari{\'{e}}}, \citenamefont {Gr{\'{e}}maud},
  \citenamefont {Delande}, \citenamefont {Szriftgiser},\ and\ \citenamefont
  {Garreau}}]{Chabe2008}%
  \BibitemOpen
  \bibfield  {author} {\bibinfo {author} {\bibfnamefont {J.}~\bibnamefont
  {Chab{\'{e}}}}, \bibinfo {author} {\bibfnamefont {G.}~\bibnamefont
  {Lemari{\'{e}}}}, \bibinfo {author} {\bibfnamefont {B.}~\bibnamefont
  {Gr{\'{e}}maud}}, \bibinfo {author} {\bibfnamefont {D.}~\bibnamefont
  {Delande}}, \bibinfo {author} {\bibfnamefont {P.}~\bibnamefont
  {Szriftgiser}},\ and\ \bibinfo {author} {\bibfnamefont {J.}~\bibnamefont
  {Garreau}},\ }\href@noop {} {\bibfield  {journal} {\bibinfo  {journal} {Phys.
  Rev. Lett.}\ }\textbf {\bibinfo {volume} {101}},\ \bibinfo {pages} {255702}
  (\bibinfo {year} {2008})}\BibitemShut {NoStop}%
\bibitem [{\citenamefont {Kondov}\ \emph {et~al.}(2011)\citenamefont {Kondov},
  \citenamefont {McGehee}, \citenamefont {Zirbel},\ and\ \citenamefont
  {DeMarco}}]{Kondov2011}%
  \BibitemOpen
  \bibfield  {author} {\bibinfo {author} {\bibfnamefont {S.~S.}\ \bibnamefont
  {Kondov}}, \bibinfo {author} {\bibfnamefont {W.~R.}\ \bibnamefont {McGehee}},
  \bibinfo {author} {\bibfnamefont {J.~J.}\ \bibnamefont {Zirbel}},\ and\
  \bibinfo {author} {\bibfnamefont {B.}~\bibnamefont {DeMarco}},\ }\href@noop
  {} {\bibfield  {journal} {\bibinfo  {journal} {Science}\ }\textbf {\bibinfo
  {volume} {334}},\ \bibinfo {pages} {66} (\bibinfo {year} {2011})}\BibitemShut
  {NoStop}%
\bibitem [{\citenamefont {Jendrzejewski}\ \emph {et~al.}(2012)\citenamefont
  {Jendrzejewski}, \citenamefont {Bernard}, \citenamefont {M{\"{u}}ller},
  \citenamefont {Cheinet}, \citenamefont {Josse}, \citenamefont {Piraud},
  \citenamefont {Pezz{\'{e}}}, \citenamefont {Sanchez-Palencia}, \citenamefont
  {Aspect},\ and\ \citenamefont {Bouyer}}]{Jendrzejewski2012}%
  \BibitemOpen
  \bibfield  {author} {\bibinfo {author} {\bibfnamefont {F.}~\bibnamefont
  {Jendrzejewski}}, \bibinfo {author} {\bibfnamefont {A.}~\bibnamefont
  {Bernard}}, \bibinfo {author} {\bibfnamefont {K.}~\bibnamefont
  {M{\"{u}}ller}}, \bibinfo {author} {\bibfnamefont {P.}~\bibnamefont
  {Cheinet}}, \bibinfo {author} {\bibfnamefont {V.}~\bibnamefont {Josse}},
  \bibinfo {author} {\bibfnamefont {M.}~\bibnamefont {Piraud}}, \bibinfo
  {author} {\bibfnamefont {L.}~\bibnamefont {Pezz{\'{e}}}}, \bibinfo {author}
  {\bibfnamefont {L.}~\bibnamefont {Sanchez-Palencia}}, \bibinfo {author}
  {\bibfnamefont {A.}~\bibnamefont {Aspect}},\ and\ \bibinfo {author}
  {\bibfnamefont {P.}~\bibnamefont {Bouyer}},\ }\href@noop {} {\bibfield
  {journal} {\bibinfo  {journal} {Nat. Phys.}\ }\textbf {\bibinfo {volume}
  {8}},\ \bibinfo {pages} {398} (\bibinfo {year} {2012})}\BibitemShut {NoStop}%
\bibitem [{\citenamefont {Skipetrov}\ and\ \citenamefont
  {Page}(2016)}]{Skipetrov2016}%
  \BibitemOpen
  \bibfield  {author} {\bibinfo {author} {\bibfnamefont {S.~E.}\ \bibnamefont
  {Skipetrov}}\ and\ \bibinfo {author} {\bibfnamefont {J.~H.}\ \bibnamefont
  {Page}},\ }\href@noop {} {\bibfield  {journal} {\bibinfo  {journal} {New J.
  Phys.}\ }\textbf {\bibinfo {volume} {18}},\ \bibinfo {pages} {21001}
  (\bibinfo {year} {2016})}\BibitemShut {NoStop}%
\bibitem [{\citenamefont {Cottier}\ \emph {et~al.}(2019)\citenamefont
  {Cottier}, \citenamefont {Cipris}, \citenamefont {Bachelard},\ and\
  \citenamefont {Kaiser}}]{Cottier2019}%
  \BibitemOpen
  \bibfield  {author} {\bibinfo {author} {\bibfnamefont {F.}~\bibnamefont
  {Cottier}}, \bibinfo {author} {\bibfnamefont {A.}~\bibnamefont {Cipris}},
  \bibinfo {author} {\bibfnamefont {R.}~\bibnamefont {Bachelard}},\ and\
  \bibinfo {author} {\bibfnamefont {R.}~\bibnamefont {Kaiser}},\ }\href@noop {}
  {\bibfield  {journal} {\bibinfo  {journal} {Phys. Rev. Lett.}\
  }\textbf {\bibinfo {volume} {123}},\ \bibinfo {pages} {083401} (\bibinfo
  {year} {2019})}\BibitemShut {NoStop}%
\bibitem [{\citenamefont {John}(1992)}]{John1992}%
  \BibitemOpen
  \bibfield  {author} {\bibinfo {author} {\bibfnamefont {S.}~\bibnamefont
  {John}},\ }\href@noop {} {\bibfield  {journal} {\bibinfo  {journal} {Phys.
  Today}\ }\textbf {\bibinfo {volume} {45}},\ \bibinfo {pages} {122} (\bibinfo
  {year} {1992})}\BibitemShut {NoStop}%
\bibitem [{\citenamefont {Skipetrov}\ and\ \citenamefont
  {Sokolov}(2014)}]{Skipetrov2014}%
  \BibitemOpen
  \bibfield  {author} {\bibinfo {author} {\bibfnamefont {S.~E.}\ \bibnamefont
  {Skipetrov}}\ and\ \bibinfo {author} {\bibfnamefont {I.~M.}\ \bibnamefont
  {Sokolov}},\ }\href@noop {} {\bibfield  {journal} {\bibinfo  {journal} {Phys.
  Rev. Lett.}\ }\textbf {\bibinfo {volume} {112}},\ \bibinfo {pages} {023905}
  (\bibinfo {year} {2014})}\BibitemShut {NoStop}%
\bibitem [{\citenamefont {Bellando}\ \emph {et~al.}(2014)\citenamefont
  {Bellando}, \citenamefont {Gero}, \citenamefont {Akkermans},\ and\
  \citenamefont {Kaiser}}]{Bellando2014}%
  \BibitemOpen
  \bibfield  {author} {\bibinfo {author} {\bibfnamefont {L.}~\bibnamefont
  {Bellando}}, \bibinfo {author} {\bibfnamefont {A.}~\bibnamefont {Gero}},
  \bibinfo {author} {\bibfnamefont {E.}~\bibnamefont {Akkermans}},\ and\
  \bibinfo {author} {\bibfnamefont {R.}~\bibnamefont {Kaiser}},\ }\href@noop {}
  {\bibfield  {journal} {\bibinfo  {journal} {Phys. Rev. A.}\ }\textbf
  {\bibinfo {volume} {90}},\ \bibinfo {pages} {063822} (\bibinfo {year}
  {2014})}\BibitemShut {NoStop}%
\bibitem [{\citenamefont {Naraghi}\ \emph {et~al.}(2015)\citenamefont
  {Naraghi}, \citenamefont {Sukhov}, \citenamefont {S{\'{a}}enz},\ and\
  \citenamefont {Dogariu}}]{Naraghi2015}%
  \BibitemOpen
  \bibfield  {author} {\bibinfo {author} {\bibfnamefont {R.~R.}\ \bibnamefont
  {Naraghi}}, \bibinfo {author} {\bibfnamefont {S.}~\bibnamefont {Sukhov}},
  \bibinfo {author} {\bibfnamefont {J.~J.}\ \bibnamefont {S{\'{a}}enz}},\ and\
  \bibinfo {author} {\bibfnamefont {A.}~\bibnamefont {Dogariu}},\ }\href@noop
  {} {\bibfield  {journal} {\bibinfo  {journal} {Phys. Rev. Lett.}\
  }\textbf {\bibinfo {volume} {115}},\ \bibinfo {pages} {203903} (\bibinfo
  {year} {2015})}\BibitemShut {NoStop}%
\bibitem [{\citenamefont {Naraghi}\ and\ \citenamefont
  {Dogariu}(2016)}]{Naraghi2016}%
  \BibitemOpen
  \bibfield  {author} {\bibinfo {author} {\bibfnamefont {R.~R.}\ \bibnamefont
  {Naraghi}}\ and\ \bibinfo {author} {\bibfnamefont {A.}~\bibnamefont
  {Dogariu}},\ }\href@noop {} {\bibfield  {journal} {\bibinfo  {journal} {Phys.
  Rev. Lett.}\ }\textbf {\bibinfo {volume} {117}},\ \bibinfo {pages} {263901}
  (\bibinfo {year} {2016})}\BibitemShut {NoStop}%
\bibitem [{\citenamefont {Escalante}\ and\ \citenamefont
  {Skipetrov}(2017)}]{Escalante2017}%
  \BibitemOpen
  \bibfield  {author} {\bibinfo {author} {\bibfnamefont {J.~M.}\ \bibnamefont
  {Escalante}}\ and\ \bibinfo {author} {\bibfnamefont {S.~E.}\ \bibnamefont
  {Skipetrov}},\ }\href@noop {} {\bibfield  {journal} {\bibinfo  {journal}
  {Ann. Phys.}\ }\textbf {\bibinfo {volume} {529}},\ \bibinfo {pages}
  {1} (\bibinfo {year} {2017})}\BibitemShut {NoStop}%
\bibitem [{\citenamefont {van Tiggelen}\ and\ \citenamefont
  {Skipetrov}(2021)}]{Tiggelen2021}%
  \BibitemOpen
  \bibfield  {author} {\bibinfo {author} {\bibfnamefont {B.~A.}\ \bibnamefont
  {van Tiggelen}}\ and\ \bibinfo {author} {\bibfnamefont {S.~E.}\ \bibnamefont
  {Skipetrov}},\ }\href@noop {} {\bibfield  {journal} {\bibinfo  {journal}
  {Phys. Rev. B}\ }\textbf {\bibinfo {volume} {103}},\ \bibinfo {pages}
  {174204} (\bibinfo {year} {2021})}\BibitemShut {NoStop}%
\bibitem [{\citenamefont {Cherroret}\ \emph {et~al.}(2016)\citenamefont
  {Cherroret}, \citenamefont {Delande},\ and\ \citenamefont
  {Van~Tiggelen}}]{Cherroret2016}%
  \BibitemOpen
  \bibfield  {author} {\bibinfo {author} {\bibfnamefont {N.}~\bibnamefont
  {Cherroret}}, \bibinfo {author} {\bibfnamefont {D.}~\bibnamefont {Delande}},\
  and\ \bibinfo {author} {\bibfnamefont {B.~A.}\ \bibnamefont {Van~Tiggelen}},\
  }\href@noop {} {\bibfield  {journal} {\bibinfo  {journal} {Phys. Rev. A}\
  }\textbf {\bibinfo {volume} {94}},\ \bibinfo {pages} {012702} (\bibinfo
  {year} {2016})}\BibitemShut {NoStop}%
\bibitem [{\citenamefont {Skipetrov}\ and\ \citenamefont
  {Sokolov}(2019)}]{Skipetrov2019}%
  \BibitemOpen
  \bibfield  {author} {\bibinfo {author} {\bibfnamefont {S.~E.}\ \bibnamefont
  {Skipetrov}}\ and\ \bibinfo {author} {\bibfnamefont {I.~M.}\ \bibnamefont
  {Sokolov}},\ }\href@noop {} {\bibfield  {journal} {\bibinfo  {journal}
  {Phys. Rev. B}\ }\textbf {\bibinfo {volume} {99}},\ \bibinfo {pages}
  {134201} (\bibinfo {year} {2019})}\BibitemShut {NoStop}%
\bibitem [{\citenamefont {Badon}\ \emph {et~al.}(2015)\citenamefont {Badon},
  \citenamefont {Lerosey}, \citenamefont {Boccara}, \citenamefont {Fink},\ and\
  \citenamefont {Aubry}}]{Badon2015}%
  \BibitemOpen
  \bibfield  {author} {\bibinfo {author} {\bibfnamefont {A.}~\bibnamefont
  {Badon}}, \bibinfo {author} {\bibfnamefont {G.}~\bibnamefont {Lerosey}},
  \bibinfo {author} {\bibfnamefont {A.~C.}\ \bibnamefont {Boccara}}, \bibinfo
  {author} {\bibfnamefont {M.}~\bibnamefont {Fink}},\ and\ \bibinfo {author}
  {\bibfnamefont {A.}~\bibnamefont {Aubry}},\ }\href@noop {} {\bibfield
  {journal} {\bibinfo  {journal} {Phys. Rev. Lett.}\ }\textbf {\bibinfo
  {volume} {114}},\ \bibinfo {pages} {023901} (\bibinfo {year}
  {2015})}\BibitemShut {NoStop}%
\bibitem [{\citenamefont {Weaver}\ and\ \citenamefont
  {Lobkis}(2001)}]{Weaver2001}%
  \BibitemOpen
  \bibfield  {author} {\bibinfo {author} {\bibfnamefont {R.~L.}\ \bibnamefont
  {Weaver}}\ and\ \bibinfo {author} {\bibfnamefont {O.~I.}\ \bibnamefont
  {Lobkis}},\ }\href@noop {} {\bibfield  {journal} {\bibinfo  {journal} {Phys.
  Rev. Lett.}\ }\textbf {\bibinfo {volume} {87}},\ \bibinfo {pages} {134301}
  (\bibinfo {year} {2001})}\BibitemShut {NoStop}%
\bibitem [{\citenamefont {Derode}\ \emph {et~al.}(2003)\citenamefont {Derode},
  \citenamefont {Larose}, \citenamefont {Campillo},\ and\ \citenamefont
  {Fink}}]{Derode2003}%
  \BibitemOpen
  \bibfield  {author} {\bibinfo {author} {\bibfnamefont {A.}~\bibnamefont
  {Derode}}, \bibinfo {author} {\bibfnamefont {E.}~\bibnamefont {Larose}},
  \bibinfo {author} {\bibfnamefont {M.}~\bibnamefont {Campillo}},\ and\
  \bibinfo {author} {\bibfnamefont {M.}~\bibnamefont {Fink}},\ }\href@noop {}
  {\bibfield  {journal} {\bibinfo  {journal} {Appl. Phys. Lett.}\ }\textbf
  {\bibinfo {volume} {83}},\ \bibinfo {pages} {3054 } (\bibinfo {year}
  {2003})}\BibitemShut {NoStop}%
\bibitem [{\citenamefont {Campillo}\ and\ \citenamefont
  {Paul}(2003)}]{Campillo2003}%
  \BibitemOpen
  \bibfield  {author} {\bibinfo {author} {\bibfnamefont {M.}~\bibnamefont
  {Campillo}}\ and\ \bibinfo {author} {\bibfnamefont {A.}~\bibnamefont
  {Paul}},\ }\href@noop {} {\bibfield  {journal} {\bibinfo  {journal}
  {Science}\ }\textbf {\bibinfo {volume} {299}},\ \bibinfo {pages} {547}
  (\bibinfo {year} {2003})}\BibitemShut {NoStop}%
\bibitem [{\citenamefont {Larose}\ \emph {et~al.}(2006)\citenamefont {Larose},
  \citenamefont {Margerin}, \citenamefont {Derode}, \citenamefont {van
  Tiggelen}, \citenamefont {Campillo}, \citenamefont {Shapiro}, \citenamefont
  {Paul}, \citenamefont {Stehly},\ and\ \citenamefont {Tanter}}]{Larose2006}%
  \BibitemOpen
  \bibfield  {author} {\bibinfo {author} {\bibfnamefont {E.}~\bibnamefont
  {Larose}}, \bibinfo {author} {\bibfnamefont {L.}~\bibnamefont {Margerin}},
  \bibinfo {author} {\bibfnamefont {A.}~\bibnamefont {Derode}}, \bibinfo
  {author} {\bibfnamefont {B.~A.}\ \bibnamefont {van Tiggelen}}, \bibinfo
  {author} {\bibfnamefont {M.}~\bibnamefont {Campillo}}, \bibinfo {author}
  {\bibfnamefont {N.}~\bibnamefont {Shapiro}}, \bibinfo {author} {\bibfnamefont
  {A.}~\bibnamefont {Paul}}, \bibinfo {author} {\bibfnamefont {L.}~\bibnamefont
  {Stehly}},\ and\ \bibinfo {author} {\bibfnamefont {M.}~\bibnamefont
  {Tanter}},\ }\href@noop {} {\bibfield  {journal} {\bibinfo  {journal}
  {Geophysics}\ }\textbf {\bibinfo {volume} {71}},\ \bibinfo {pages} {SI11}
  (\bibinfo {year} {2006})}\BibitemShut {NoStop}%
\bibitem [{\citenamefont {Badon}\ \emph {et~al.}(2016)\citenamefont {Badon},
  \citenamefont {Li}, \citenamefont {Lerosey}, \citenamefont {Boccara},
  \citenamefont {Fink},\ and\ \citenamefont {Aubry}}]{Badon2016}%
  \BibitemOpen
  \bibfield  {author} {\bibinfo {author} {\bibfnamefont {A.}~\bibnamefont
  {Badon}}, \bibinfo {author} {\bibfnamefont {D.}~\bibnamefont {Li}}, \bibinfo
  {author} {\bibfnamefont {G.}~\bibnamefont {Lerosey}}, \bibinfo {author}
  {\bibfnamefont {A.~C.}\ \bibnamefont {Boccara}}, \bibinfo {author}
  {\bibfnamefont {M.}~\bibnamefont {Fink}},\ and\ \bibinfo {author}
  {\bibfnamefont {A.}~\bibnamefont {Aubry}},\ }\href@noop {} {\bibfield
  {journal} {\bibinfo  {journal} {Optica}\ }\textbf {\bibinfo {volume} {3}},\
  \bibinfo {pages} {11} (\bibinfo {year} {2016})}\BibitemShut {NoStop}%
\bibitem [{\citenamefont {Page}\ \emph {et~al.}(1995)\citenamefont {Page},
  \citenamefont {Schriemer}, \citenamefont {Bailey},\ and\ \citenamefont
  {Weitz}}]{Page1995}%
  \BibitemOpen
  \bibfield  {author} {\bibinfo {author} {\bibfnamefont {J.~H.}\ \bibnamefont
  {Page}}, \bibinfo {author} {\bibfnamefont {H.~P.}\ \bibnamefont {Schriemer}},
  \bibinfo {author} {\bibfnamefont {A.~E.}\ \bibnamefont {Bailey}},\ and\
  \bibinfo {author} {\bibfnamefont {D.~A.}\ \bibnamefont {Weitz}},\ }\href@noop
  {} {\bibfield  {journal} {\bibinfo  {journal} {Phys. Rev. E}\ }\textbf
  {\bibinfo {volume} {52}},\ \bibinfo {pages} {3106} (\bibinfo {year}
  {1995})}\BibitemShut {NoStop}%
\bibitem [{\citenamefont {B\"uhrer}(2012)}]{BuhrerPhD2012}%
  \BibitemOpen
  \bibfield  {author} {\bibinfo {author} {\bibfnamefont {W.}~\bibnamefont
  {B\"uhrer}},\ }{\bibinfo {title} {{Anderson Localization of Light in
  the Presence of Non-linear Effects}}},\ \href@noop {} {Ph.D. thesis},\
  \bibinfo  {school} {Universit\"at Konstanz} (\bibinfo {year}
  {2012})\BibitemShut {NoStop}%
\bibitem [{\citenamefont {Abrahams}\ \emph {et~al.}(1979)\citenamefont
  {Abrahams}, \citenamefont {Anderson}, \citenamefont {Licciardello},\ and\
  \citenamefont {Ramakrishnan}}]{Abrahams1979}%
  \BibitemOpen
  \bibfield  {author} {\bibinfo {author} {\bibfnamefont {E.}~\bibnamefont
  {Abrahams}}, \bibinfo {author} {\bibfnamefont {P.~W.}\ \bibnamefont
  {Anderson}}, \bibinfo {author} {\bibfnamefont {D.~C.}\ \bibnamefont
  {Licciardello}},\ and\ \bibinfo {author} {\bibfnamefont {T.~V.}\ \bibnamefont
  {Ramakrishnan}},\ }\href@noop {} {\bibfield  {journal} {\bibinfo  {journal}
  {Phys. Rev. Lett.}\ }\textbf {\bibinfo {volume} {42}},\ \bibinfo {pages}
  {673} (\bibinfo {year} {1979})}\BibitemShut {NoStop}%
\bibitem [{\citenamefont {Sperling}(2015)}]{SperlingPhD2015}%
  \BibitemOpen
  \bibfield  {author} {\bibinfo {author} {\bibfnamefont {T.}~\bibnamefont
  {Sperling}},\ }{\bibinfo {title} {{The experimental search for Anderson
  localisation of light in three-dimensions}}},\ \href@noop {} {Ph.D. thesis},\
  \bibinfo  {school} {Universit\"at Konstanz} (\bibinfo {year}
  {2015})\BibitemShut {NoStop}%
\bibitem [{\citenamefont {Schertel}\ \emph {et~al.}(2019)\citenamefont
  {Schertel}, \citenamefont {Wimmer}, \citenamefont {Besirske}, \citenamefont
  {Aegerter}, \citenamefont {Maret}, \citenamefont {Polarz},\ and\
  \citenamefont {Aubry}}]{Schertel2019}%
  \BibitemOpen
  \bibfield  {author} {\bibinfo {author} {\bibfnamefont {L.}~\bibnamefont
  {Schertel}}, \bibinfo {author} {\bibfnamefont {I.}~\bibnamefont {Wimmer}},
  \bibinfo {author} {\bibfnamefont {P.}~\bibnamefont {Besirske}}, \bibinfo
  {author} {\bibfnamefont {C.~M.}\ \bibnamefont {Aegerter}}, \bibinfo {author}
  {\bibfnamefont {G.}~\bibnamefont {Maret}}, \bibinfo {author} {\bibfnamefont
  {S.}~\bibnamefont {Polarz}},\ and\ \bibinfo {author} {\bibfnamefont {G.~J.}\
  \bibnamefont {Aubry}},\ }\href@noop {} {\bibfield  {journal} {\bibinfo
  {journal} {Phys. Rev. M}\ }\textbf {\bibinfo {volume} {3}},\ \bibinfo {pages}
  {015203} (\bibinfo {year} {2019})}\BibitemShut {NoStop}%
\bibitem [{\citenamefont {Bevington}\ and\ \citenamefont
  {Robinson}(1992)}]{Bevington2002}%
  \BibitemOpen
  \bibfield  {author} {\bibinfo {author} {\bibfnamefont {P.~R.}\ \bibnamefont
  {Bevington}}\ and\ \bibinfo {author} {\bibfnamefont {D.~K.}\ \bibnamefont
  {Robinson}},\ }\href@noop {} { {\bibinfo {title} {{ Data Reduction and
  Error Analysis for the Physical Sciences}}}},\ \bibinfo {edition} {2nd}\ ed.\
  (\bibinfo  {publisher} {McGraw-Hill, New York},\ \bibinfo {year}
  {1992})\BibitemShut {NoStop}%
\bibitem [{\citenamefont {Aubry}\ and\ \citenamefont
  {Derode}(2007)}]{Aubry2007}%
  \BibitemOpen
  \bibfield  {author} {\bibinfo {author} {\bibfnamefont {A.}~\bibnamefont
  {Aubry}}\ and\ \bibinfo {author} {\bibfnamefont {A.}~\bibnamefont {Derode}},\
  }\href@noop {} {\bibfield  {journal} {\bibinfo  {journal} {Phys. Rev. E}\
  }\textbf {\bibinfo {volume} {75}},\ \bibinfo {pages} {026602} (\bibinfo
  {year} {2007})}\BibitemShut {NoStop}%
\bibitem [{\citenamefont {Patterson}\ \emph {et~al.}(1989)\citenamefont
  {Patterson}, \citenamefont {Chance},\ and\ \citenamefont
  {Wilson}}]{Patterson1989}%
  \BibitemOpen
  \bibfield  {author} {\bibinfo {author} {\bibfnamefont {M.~S.}\ \bibnamefont
  {Patterson}}, \bibinfo {author} {\bibfnamefont {B.}~\bibnamefont {Chance}},\
  and\ \bibinfo {author} {\bibfnamefont {B.~C.}\ \bibnamefont {Wilson}},\
  }\href@noop {} {\bibfield  {journal} {\bibinfo  {journal} {Appl. Opt.}\
  }\textbf {\bibinfo {volume} {28}},\ \bibinfo {pages} {2331} (\bibinfo {year}
  {1989})}\BibitemShut {NoStop}%
\bibitem [{\citenamefont {Lemari{\'{e}}}\ \emph {et~al.}(2010)\citenamefont
  {Lemari{\'{e}}}, \citenamefont {Lignier}, \citenamefont {Delande},
  \citenamefont {Szriftgiser},\ and\ \citenamefont {Garreau}}]{Lemarie2010}%
  \BibitemOpen
  \bibfield  {author} {\bibinfo {author} {\bibfnamefont {G.}~\bibnamefont
  {Lemari{\'{e}}}}, \bibinfo {author} {\bibfnamefont {H.}~\bibnamefont
  {Lignier}}, \bibinfo {author} {\bibfnamefont {D.}~\bibnamefont {Delande}},
  \bibinfo {author} {\bibfnamefont {P.}~\bibnamefont {Szriftgiser}},\ and\
  \bibinfo {author} {\bibfnamefont {J.~C.}\ \bibnamefont {Garreau}},\
  }\href@noop {} {\bibfield  {journal} {\bibinfo  {journal} {Phys. Rev. Lett.}\
  }\textbf {\bibinfo {volume} {105}},\ \bibinfo {pages} {090601} (\bibinfo
  {year} {2010})}\BibitemShut {NoStop}%
\bibitem [{\citenamefont {Cherroret}\ \emph {et~al.}(2010)\citenamefont
  {Cherroret}, \citenamefont {Skipetrov},\ and\ \citenamefont {van
  Tiggelen}}]{Cherroret2010}%
  \BibitemOpen
  \bibfield  {author} {\bibinfo {author} {\bibfnamefont {N.}~\bibnamefont
  {Cherroret}}, \bibinfo {author} {\bibfnamefont {S.~E.}\ \bibnamefont
  {Skipetrov}},\ and\ \bibinfo {author} {\bibfnamefont {B.~A.}\ \bibnamefont
  {van Tiggelen}},\ }\href@noop {} {\bibfield  {journal} {\bibinfo  {journal}
  {Phys. Rev. E.}\ }\textbf {\bibinfo {volume} {82}},\ \bibinfo {pages}
  {056603} (\bibinfo {year} {2010})}\BibitemShut {NoStop}%
\bibitem [{\citenamefont {Van~Albada}\ and\ \citenamefont
  {Lagendijk}(1985)}]{Albada1985}%
  \BibitemOpen
  \bibfield  {author} {\bibinfo {author} {\bibfnamefont {M.~P.}\ \bibnamefont
  {Van~Albada}}\ and\ \bibinfo {author} {\bibfnamefont {A.}~\bibnamefont
  {Lagendijk}},\ }\href {https://doi.org/10.1103/physrevlett.55.2692}
  {\bibfield  {journal} {\bibinfo  {journal} {Phys. Rev. Lett.}\ }\textbf
  {\bibinfo {volume} {55}},\ \bibinfo {pages} {2692} (\bibinfo {year}
  {1985})}\BibitemShut {NoStop}%
\bibitem [{\citenamefont {Wolf}\ and\ \citenamefont {Maret}(1985)}]{Wolf1985}%
  \BibitemOpen
  \bibfield  {author} {\bibinfo {author} {\bibfnamefont {P.-E.}\ \bibnamefont
  {Wolf}}\ and\ \bibinfo {author} {\bibfnamefont {G.}~\bibnamefont {Maret}},\
  }\href {https://doi.org/10.1103/physrevlett.55.2696} {\bibfield  {journal}
  {\bibinfo  {journal} {Phys. Rev. Lett.}\ }\textbf {\bibinfo {volume} {55}},\
  \bibinfo {pages} {2696} (\bibinfo {year} {1985})}\BibitemShut {NoStop}%
\bibitem [{\citenamefont {Vreeker}\ \emph {et~al.}(1988)\citenamefont
  {Vreeker}, \citenamefont {{van Albada}}, \citenamefont {Sprik},\ and\
  \citenamefont {Lagendijk}}]{Vreeker1988}%
  \BibitemOpen
  \bibfield  {author} {\bibinfo {author} {\bibfnamefont {R.}~\bibnamefont
  {Vreeker}}, \bibinfo {author} {\bibfnamefont {M.~P.}\ \bibnamefont {{van
  Albada}}}, \bibinfo {author} {\bibfnamefont {R.}~\bibnamefont {Sprik}},\ and\
  \bibinfo {author} {\bibfnamefont {A.}~\bibnamefont {Lagendijk}},\ }\href@noop
  {} {\bibfield  {journal} {\bibinfo  {journal} {Phys. Lett. A}\ }\textbf
  {\bibinfo {volume} {132}},\ \bibinfo {pages} {51} (\bibinfo {year}
  {1988})}\BibitemShut {NoStop}%
\bibitem [{\citenamefont {Tourin}\ \emph {et~al.}(1997)\citenamefont {Tourin},
  \citenamefont {Derode}, \citenamefont {Roux}, \citenamefont {van Tiggelen},\
  and\ \citenamefont {Fink}}]{Tourin1997}%
  \BibitemOpen
  \bibfield  {author} {\bibinfo {author} {\bibfnamefont {A.}~\bibnamefont
  {Tourin}}, \bibinfo {author} {\bibfnamefont {A.}~\bibnamefont {Derode}},
  \bibinfo {author} {\bibfnamefont {P.}~\bibnamefont {Roux}}, \bibinfo {author}
  {\bibfnamefont {B.~A.}\ \bibnamefont {van Tiggelen}},\ and\ \bibinfo {author}
  {\bibfnamefont {M.}~\bibnamefont {Fink}},\ }\href
  {https://doi.org/10.1103/physrevlett.79.3637} {\bibfield  {journal} {\bibinfo
   {journal} {Phys. Rev. Lett.}\ }\textbf {\bibinfo {volume} {79}},\
  \bibinfo {pages} {3637} (\bibinfo {year} {1997})}\BibitemShut {NoStop}%
\bibitem [{\citenamefont {Hainaut}\ \emph {et~al.}(2017)\citenamefont
  {Hainaut}, \citenamefont {Manai}, \citenamefont {Chicireanu}, \citenamefont
  {Cl{\'{e}}ment}, \citenamefont {Zemmouri}, \citenamefont {Garreau},
  \citenamefont {Szriftgiser}, \citenamefont {Lemari{\'{e}}}, \citenamefont
  {Cherroret},\ and\ \citenamefont {Delande}}]{Hainaut2017}%
  \BibitemOpen
  \bibfield  {author} {\bibinfo {author} {\bibfnamefont {C.}~\bibnamefont
  {Hainaut}}, \bibinfo {author} {\bibfnamefont {I.}~\bibnamefont {Manai}},
  \bibinfo {author} {\bibfnamefont {R.}~\bibnamefont {Chicireanu}}, \bibinfo
  {author} {\bibfnamefont {J.-F.}\ \bibnamefont {Cl{\'{e}}ment}}, \bibinfo
  {author} {\bibfnamefont {S.}~\bibnamefont {Zemmouri}}, \bibinfo {author}
  {\bibfnamefont {J.-C.}\ \bibnamefont {Garreau}}, \bibinfo {author}
  {\bibfnamefont {P.}~\bibnamefont {Szriftgiser}}, \bibinfo {author}
  {\bibfnamefont {G.}~\bibnamefont {Lemari{\'{e}}}}, \bibinfo {author}
  {\bibfnamefont {N.}~\bibnamefont {Cherroret}},\ and\ \bibinfo {author}
  {\bibfnamefont {D.}~\bibnamefont {Delande}},\ }\href@noop {} {\bibfield
  {journal} {\bibinfo  {journal} {Phys. Rev. Lett.}\ }\textbf {\bibinfo
  {volume} {118}},\ \bibinfo {pages} {184101} (\bibinfo {year}
  {2017})}\BibitemShut {NoStop}%
\bibitem [{\citenamefont {Margerin}\ \emph {et~al.}(2001)\citenamefont
  {Margerin}, \citenamefont {Campillo},\ and\ \citenamefont {van
  Tiggelen}}]{Margerin2001}%
  \BibitemOpen
  \bibfield  {author} {\bibinfo {author} {\bibfnamefont {L.}~\bibnamefont
  {Margerin}}, \bibinfo {author} {\bibfnamefont {M.}~\bibnamefont {Campillo}},\
  and\ \bibinfo {author} {\bibfnamefont {B.~A.}\ \bibnamefont {van Tiggelen}},\
  }\href@noop {} {\bibfield  {journal} {\bibinfo  {journal} {Geophys. J. Int.}\
  }\textbf {\bibinfo {volume} {145}},\ \bibinfo {pages} {593} (\bibinfo {year}
  {2001})}\BibitemShut {NoStop}%
\bibitem [{\citenamefont {Larose}\ \emph {et~al.}(2004)\citenamefont {Larose},
  \citenamefont {Margerin}, \citenamefont {van Tiggelen},\ and\ \citenamefont
  {Campillo}}]{Larose2004}%
  \BibitemOpen
  \bibfield  {author} {\bibinfo {author} {\bibfnamefont {E.}~\bibnamefont
  {Larose}}, \bibinfo {author} {\bibfnamefont {L.}~\bibnamefont {Margerin}},
  \bibinfo {author} {\bibfnamefont {B.~A.}\ \bibnamefont {van Tiggelen}},\ and\
  \bibinfo {author} {\bibfnamefont {M.}~\bibnamefont {Campillo}},\ }\href
  {https://doi.org/10.1103/physrevlett.93.048501} {\bibfield  {journal}
  {\bibinfo  {journal} {Phys. Rev. Lett.}\ }\textbf {\bibinfo {volume} {93}},\
  \bibinfo {pages} {048501} (\bibinfo {year} {2004})}\BibitemShut {NoStop}%
\bibitem [{\citenamefont {Cherroret}\ \emph {et~al.}(2021)\citenamefont
  {Cherroret}, \citenamefont {Scoquart},\ and\ \citenamefont
  {Delande}}]{Cherroret2021}%
  \BibitemOpen
  \bibfield  {author} {\bibinfo {author} {\bibfnamefont {N.}~\bibnamefont
  {Cherroret}}, \bibinfo {author} {\bibfnamefont {T.}~\bibnamefont
  {Scoquart}},\ and\ \bibinfo {author} {\bibfnamefont {D.}~\bibnamefont
  {Delande}},\ }\href {https://doi.org/10.1016/j.aop.2021.168543} {\bibfield
  {journal} {\bibinfo  {journal} {Ann. Phys.}\ }\textbf {\bibinfo
  {volume} {435}},\ \bibinfo {pages} {168543} (\bibinfo {year}
  {2021})}\BibitemShut {NoStop}%
\bibitem [{\citenamefont {Cobus}\ \emph {et~al.}(2018)\citenamefont {Cobus},
  \citenamefont {Hildebrand}, \citenamefont {Skipetrov}, \citenamefont {van
  Tiggelen},\ and\ \citenamefont {Page}}]{Cobus2018}%
  \BibitemOpen
  \bibfield  {author} {\bibinfo {author} {\bibfnamefont {L.~A.}\ \bibnamefont
  {Cobus}}, \bibinfo {author} {\bibfnamefont {W.~K.}\ \bibnamefont
  {Hildebrand}}, \bibinfo {author} {\bibfnamefont {S.~E.}\ \bibnamefont
  {Skipetrov}}, \bibinfo {author} {\bibfnamefont {B.~A.}\ \bibnamefont {van
  Tiggelen}},\ and\ \bibinfo {author} {\bibfnamefont {J.~H.}\ \bibnamefont
  {Page}},\ }\href@noop {} {\bibfield  {journal} {\bibinfo  {journal} {Phys.
  Rev. B.}\ }\textbf {\bibinfo {volume} {98}},\ \bibinfo {pages} {214201}
  (\bibinfo {year} {2018})}\BibitemShut {NoStop}%
\bibitem [{\citenamefont {Vollhardt}\ and\ \citenamefont
  {W{\"{o}}lfle}(1982)}]{Vollhardt1982}%
  \BibitemOpen
  \bibfield  {author} {\bibinfo {author} {\bibfnamefont {D.}~\bibnamefont
  {Vollhardt}}\ and\ \bibinfo {author} {\bibfnamefont {P.}~\bibnamefont
  {W{\"{o}}lfle}},\ }\href@noop {} {\bibfield  {journal} {\bibinfo  {journal}
  {Phys. Rev. Lett.}\ }\textbf {\bibinfo {volume} {48}},\ \bibinfo {pages}
  {699} (\bibinfo {year} {1982})}\BibitemShut {NoStop}%
\bibitem [{\citenamefont {Skipetrov}\ and\ \citenamefont {van
  Tiggelen}(2006)}]{Skipetrov2006}%
  \BibitemOpen
  \bibfield  {author} {\bibinfo {author} {\bibfnamefont {S.~E.}\ \bibnamefont
  {Skipetrov}}\ and\ \bibinfo {author} {\bibfnamefont {B.~A.}\ \bibnamefont
  {van Tiggelen}},\ }\href {https://doi.org/10.1103/PhysRevLett.96.043902}
  {\bibfield  {journal} {\bibinfo  {journal} {Phys. Rev. Lett.}\ }\textbf
  {\bibinfo {volume} {96}},\ \bibinfo {pages} {043902} (\bibinfo {year}
  {2006})} \BibitemShut
  {NoStop}%
\bibitem [{\citenamefont {Aubry}\ \emph {et~al.}(2014)\citenamefont {Aubry},
  \citenamefont {Cobus}, \citenamefont {Skipetrov}, \citenamefont {van
  Tiggelen}, \citenamefont {Derode},\ and\ \citenamefont {Page}}]{Aubry2014}%
  \BibitemOpen
  \bibfield  {author} {\bibinfo {author} {\bibfnamefont {A.}~\bibnamefont
  {Aubry}}, \bibinfo {author} {\bibfnamefont {L.~A.}\ \bibnamefont {Cobus}},
  \bibinfo {author} {\bibfnamefont {S.~E.}\ \bibnamefont {Skipetrov}}, \bibinfo
  {author} {\bibfnamefont {B.~A.}\ \bibnamefont {van Tiggelen}}, \bibinfo
  {author} {\bibfnamefont {A.}~\bibnamefont {Derode}},\ and\ \bibinfo {author}
  {\bibfnamefont {J.~H.}\ \bibnamefont {Page}},\ }\href@noop {} {\bibfield
  {journal} {\bibinfo  {journal} {Phys. Rev. Lett.}\ }\textbf {\bibinfo
  {volume} {112}},\ \bibinfo {pages} {043903} (\bibinfo {year}
  {2014})}\BibitemShut {NoStop}%
\bibitem [{\citenamefont {Douglass}\ \emph {et~al.}(2011)\citenamefont
  {Douglass}, \citenamefont {John}, \citenamefont {Suezaki}, \citenamefont
  {Ozin},\ and\ \citenamefont {Dogariu}}]{Douglass2011}%
  \BibitemOpen
  \bibfield  {author} {\bibinfo {author} {\bibfnamefont {K.~M.}\ \bibnamefont
  {Douglass}}, \bibinfo {author} {\bibfnamefont {S.}~\bibnamefont {John}},
  \bibinfo {author} {\bibfnamefont {T.}~\bibnamefont {Suezaki}}, \bibinfo
  {author} {\bibfnamefont {G.~A.}\ \bibnamefont {Ozin}},\ and\ \bibinfo
  {author} {\bibfnamefont {A.}~\bibnamefont {Dogariu}},\ }\href@noop {}
  {\bibfield  {journal} {\bibinfo  {journal} {Opt. Express}\ }\textbf {\bibinfo
  {volume} {19}},\ \bibinfo {pages} {25320} (\bibinfo {year}
  {2011})}\BibitemShut {NoStop}%
\bibitem [{foo()}]{footnote}%
  \BibitemOpen
  \href@noop {} {}\bibinfo {note} {The temporal scaling laws of the return
  probability differ from the ones considered in a previous acoustics
  work~\cite{Aubry2014}. While the previous study considered the far-field
  reflectance of the medium, \textit{i.e} an energy flux
  density~\cite{Patterson1989}, here we probe the energy density at the sample
  surface.}\BibitemShut {Stop}%
\bibitem [{\citenamefont {Akkermans}\ and\ \citenamefont
  {Montambaux}(2007)}]{Akkermans2007}%
  \BibitemOpen
  \bibfield  {author} {\bibinfo {author} {\bibfnamefont {E.}~\bibnamefont
  {Akkermans}}\ and\ \bibinfo {author} {\bibfnamefont {G.}~\bibnamefont
  {Montambaux}},\ }\href@noop {} { {\bibinfo {title} {Mesoscopic Physics
  of Electrons and Photons}}}\ (\bibinfo  {publisher} {Cambridge University
  Press},\ \bibinfo {address} {Cambridge},\ \bibinfo {year} {2007})\BibitemShut
  {NoStop}%
\bibitem [{\citenamefont {Skipetrov}\ and\ \citenamefont
  {Sinha}(2018)}]{Skipetrov2018}%
  \BibitemOpen
  \bibfield  {author} {\bibinfo {author} {\bibfnamefont {S.~E.}\ \bibnamefont
  {Skipetrov}}\ and\ \bibinfo {author} {\bibfnamefont {A.}~\bibnamefont
  {Sinha}},\ }\href@noop {} {\bibfield  {journal} {\bibinfo  {journal}
  {Phys. Rev. B}\ }\textbf {\bibinfo {volume} {97}},\ \bibinfo {pages}
  {104202} (\bibinfo {year} {2018})}\BibitemShut {NoStop}%
\bibitem [{\citenamefont {Akridas-Morel}\ \emph {et~al.}(2019)\citenamefont
  {Akridas-Morel}, \citenamefont {Cherroret},\ and\ \citenamefont
  {Delande}}]{Akridas-Morel2019}%
  \BibitemOpen
  \bibfield  {author} {\bibinfo {author} {\bibfnamefont {P.}~\bibnamefont
  {Akridas-Morel}}, \bibinfo {author} {\bibfnamefont {N.}~\bibnamefont
  {Cherroret}},\ and\ \bibinfo {author} {\bibfnamefont {D.}~\bibnamefont
  {Delande}},\ }\href@noop {} {\bibfield  {journal} {\bibinfo  {journal}
  {Phys. Rev. A}\ }\textbf {\bibinfo {volume} {100}},\ \bibinfo {pages}
  {043612} (\bibinfo {year} {2019})}\BibitemShut {NoStop}%
\bibitem [{\citenamefont {Andreoli}\ \emph {et~al.}(2021)\citenamefont
  {Andreoli}, \citenamefont {Gullans}, \citenamefont {High}, \citenamefont
  {Browaeys},\ and\ \citenamefont {Chang}}]{Andreoli2021}%
  \BibitemOpen
  \bibfield  {author} {\bibinfo {author} {\bibfnamefont {F.}~\bibnamefont
  {Andreoli}}, \bibinfo {author} {\bibfnamefont {M.~J.}\ \bibnamefont
  {Gullans}}, \bibinfo {author} {\bibfnamefont {A.~A.}\ \bibnamefont {High}},
  \bibinfo {author} {\bibfnamefont {A.}~\bibnamefont {Browaeys}},\ and\
  \bibinfo {author} {\bibfnamefont {D.~E.}\ \bibnamefont {Chang}},\ }\href@noop
  {} {\bibfield  {journal} {\bibinfo  {journal} {Phys. Rev. X}\ }\textbf
  {\bibinfo {volume} {11}},\ \bibinfo {pages} {011026} (\bibinfo {year}
  {2021})}\BibitemShut {NoStop}%
\bibitem [{\citenamefont {Haberko}\ \emph {et~al.}(2020)\citenamefont
  {Haberko}, \citenamefont {Froufe-P{\'{e}}rez},\ and\ \citenamefont
  {Scheffold}}]{Haberko2020}%
  \BibitemOpen
  \bibfield  {author} {\bibinfo {author} {\bibfnamefont {J.}~\bibnamefont
  {Haberko}}, \bibinfo {author} {\bibfnamefont {L.~S.}\ \bibnamefont
  {Froufe-P{\'{e}}rez}},\ and\ \bibinfo {author} {\bibfnamefont
  {F.}~\bibnamefont {Scheffold}},\ }\bibfield  {journal} {\bibinfo  {journal}
  {Nat. Commun.}\ }\textbf {\bibinfo {volume} {11}},\ \bibinfo {pages} {4867}
  (\bibinfo {year} {2020})\BibitemShut {NoStop}%
\bibitem [{\citenamefont {Froufe-P{\'{e}}rez}\ \emph
  {et~al.}(2017)\citenamefont {Froufe-P{\'{e}}rez}, \citenamefont {Engel},
  \citenamefont {S{\'{a}}enz},\ and\ \citenamefont
  {Scheffold}}]{FroufePerez2017}%
  \BibitemOpen
  \bibfield  {author} {\bibinfo {author} {\bibfnamefont {L.~S.}\ \bibnamefont
  {Froufe-P{\'{e}}rez}}, \bibinfo {author} {\bibfnamefont {M.}~\bibnamefont
  {Engel}}, \bibinfo {author} {\bibfnamefont {J.~J.}\ \bibnamefont
  {S{\'{a}}enz}},\ and\ \bibinfo {author} {\bibfnamefont {F.}~\bibnamefont
  {Scheffold}},\ }\href {https://doi.org/10.1073/pnas.1705130114} {\bibfield
  {journal} {\bibinfo  {journal} {Proc. Natl. Acad. Sci. USA}\ }\textbf {\bibinfo {volume} {114}},\ \bibinfo {pages} {9570}
  (\bibinfo {year} {2017})}\BibitemShut {NoStop}%
\bibitem [{\citenamefont {Ghosh}\ \emph {et~al.}(2015)\citenamefont {Ghosh},
  \citenamefont {Delande}, \citenamefont {Miniatura},\ and\ \citenamefont
  {Cherroret}}]{Ghosh2015}%
  \BibitemOpen
  \bibfield  {author} {\bibinfo {author} {\bibfnamefont {S.}~\bibnamefont
  {Ghosh}}, \bibinfo {author} {\bibfnamefont {D.}~\bibnamefont {Delande}},
  \bibinfo {author} {\bibfnamefont {C.}~\bibnamefont {Miniatura}},\ and\
  \bibinfo {author} {\bibfnamefont {N.}~\bibnamefont {Cherroret}},\ }\href@noop
  {} {\bibfield  {journal} {\bibinfo  {journal} {Phys. Rev. Lett.}\
  }\textbf {\bibinfo {volume} {115}},\ \bibinfo {pages} {200602} (\bibinfo
  {year} {2015})}\BibitemShut {NoStop}%
\bibitem [{\citenamefont {Vynck}\ \emph {et~al.}(2021)\citenamefont {Vynck},
  \citenamefont {Pierrat}, \citenamefont {Carminati}, \citenamefont
  {Froufe-Pérez}, \citenamefont {Scheffold}, \citenamefont {Sapienza},
  \citenamefont {Vignolini},\ and\ \citenamefont {Sáenz}}]{Vynck2021}%
  \BibitemOpen
  \bibfield  {author} {\bibinfo {author} {\bibfnamefont {K.}~\bibnamefont
  {Vynck}}, \bibinfo {author} {\bibfnamefont {R.}~\bibnamefont {Pierrat}},
  \bibinfo {author} {\bibfnamefont {R.}~\bibnamefont {Carminati}}, \bibinfo
  {author} {\bibfnamefont {L.~S.}\ \bibnamefont {Froufe-Pérez}}, \bibinfo
  {author} {\bibfnamefont {F.}~\bibnamefont {Scheffold}}, \bibinfo {author}
  {\bibfnamefont {R.}~\bibnamefont {Sapienza}}, \bibinfo {author}
  {\bibfnamefont {S.}~\bibnamefont {Vignolini}},\ and\ \bibinfo {author}
  {\bibfnamefont {J.-J.}\ \bibnamefont {Sáenz}},\ }\href@noop {} {\bibfield
  {journal} {\bibinfo  {journal} {arXiv:2106.13892}\ } (\bibinfo {year}
  {2021})}\BibitemShut {NoStop}%
\bibitem [{\citenamefont {Goodman}(1996)}]{Goodman}%
  \BibitemOpen
  \bibfield  {author} {\bibinfo {author} {\bibfnamefont {J.~W.}\ \bibnamefont
  {Goodman}},\ }\href@noop {} { {\bibinfo {title} {Introduction to Fourier
  Optics}}}\ (\bibinfo  {publisher} {Mc Graw Hill, New York},\ \bibinfo {year}
  {1996})\BibitemShut {NoStop}%
\bibitem [{\citenamefont {Weaver}\ and\ \citenamefont
  {Lobkis}(2004)}]{Weaver2004}%
  \BibitemOpen
  \bibfield  {author} {\bibinfo {author} {\bibfnamefont {R.~L.}\ \bibnamefont
  {Weaver}}\ and\ \bibinfo {author} {\bibfnamefont {O.~I.}\ \bibnamefont
  {Lobkis}},\ }\href@noop {} {\bibfield  {journal} {\bibinfo  {journal} {J.
  Acoust. Soc. Am.}\ }\textbf {\bibinfo {volume} {116}},\ \bibinfo {pages}
  {2731} (\bibinfo {year} {2004})}\BibitemShut {NoStop}%
\bibitem [{\citenamefont {Barolle}\ \emph {et~al.}(2021)\citenamefont
  {Barolle}, \citenamefont {Scholler}, \citenamefont {Mec\^e}, \citenamefont
  {Chassot}, \citenamefont {Groux}, \citenamefont {Fink}, \citenamefont
  {Boccara},\ and\ \citenamefont {Aubry}}]{Barolle2021}%
  \BibitemOpen
  \bibfield  {author} {\bibinfo {author} {\bibfnamefont {V.}~\bibnamefont
  {Barolle}}, \bibinfo {author} {\bibfnamefont {J.}~\bibnamefont {Scholler}},
  \bibinfo {author} {\bibfnamefont {P.}~\bibnamefont {Mec\^e}}, \bibinfo
  {author} {\bibfnamefont {J.-M.}\ \bibnamefont {Chassot}}, \bibinfo {author}
  {\bibfnamefont {K.}~\bibnamefont {Groux}}, \bibinfo {author} {\bibfnamefont
  {M.}~\bibnamefont {Fink}}, \bibinfo {author} {\bibfnamefont {A.~C.}\
  \bibnamefont {Boccara}},\ and\ \bibinfo {author} {\bibfnamefont
  {A.}~\bibnamefont {Aubry}},\ }\href@noop {} {\bibfield  {journal} {\bibinfo
  {journal} {Opt. Express}\ }\textbf {\bibinfo {volume} {29}},\ \bibinfo
  {pages} {22044} (\bibinfo {year} {2021})}\BibitemShut {NoStop}%
\end{thebibliography}
%

\end{document}